\documentclass[a4paper, journal]{IEEEtran}
\usepackage{cite}
\usepackage{latexsym}
\usepackage{textgreek}
\usepackage{etoolbox}
\makeatletter
\patchcmd{\@makecaption}
  {\scshape}
  {}
  {}
  {}
\makeatletter
\patchcmd{\@makecaption}
  {\\}
  {.\ }
  {}
  {}
\makeatother
\def\tablename{Table}

\usepackage{tcolorbox}
\usepackage[table]{xcolor}
\definecolor{tableheader}{HTML}{343434}
\usepackage{color}
\usepackage{fontawesome}
\usepackage{amsthm}
\usepackage{tabularx}
\usepackage{placeins}
\usepackage{amsxtra}
\usepackage[mathscr]{euscript}
\usepackage{graphicx}
\graphicspath{{./fig/}}
\usepackage{multirow}
\usepackage{pifont}

\usepackage{rotating, graphicx}
\usepackage{array,multirow,graphicx}
\usepackage{amsmath}
\usepackage{tikz}
\usepackage{amssymb}
\usepackage{lipsum}
\usepackage{lipsum, multicol}
\usepackage{adjustbox}
\usepackage{algorithm}
\usepackage{algpseudocode}
\usepackage{upgreek}
\usepackage{mathtools}
\usepackage{float}
\usepackage{xcolor}
\usepackage[caption = false]{subfig}
\usepackage{acronym}
\usepackage{orcidlink}
\usepackage{cuted} 
\usepackage{fancyhdr}
\hypersetup{hidelinks}
\usepackage{colortbl,booktabs}
\usepackage{svg}
\usepackage{supertabular}
\usepackage{comment}
\usepackage{alltt}
\usepackage{enumitem}
\usepackage{mathtools}
\usepackage{float}
\usepackage[caption = false]{subfig}
\usepackage{threeparttable}
\usepackage[utf8]{inputenc}
\usepackage[english]{babel}

\begin{document}

    
\begin{figure*}[ht] 
\centering
\begin{tcolorbox}[
colframe=black!75!black, 
colback=darkgray!5!white,   
coltitle=white,         
colbacktitle=darkgray!75!black, 
fonttitle=\bfseries, 
title= ,  
width=\textwidth,       
boxrule=0.3mm,            
sharp corners,          
left=1mm, right=1mm,    
top=1mm, bottom=1mm     
]
\textbf{Authors’ Note:} This is a preprint version of the article currently under review. We welcome constructive feedback and comments to further enhance the quality and clarity of this work. If you have any suggestions, insights, or critiques, please feel free to contact the corresponding author, Mohammad Asif Habibi\orcidlink{0000-0001-9874-0047}, at [\textcolor{blue}{mohammad\_asif.habibi@dfki.de}]. We sincerely appreciate all feedback, whether minor, critical, or substantial.
\\ \\ 

\textbf{Disclaimer:} This article has been submitted to the IEEE for possible publication. Copyright may be transferred without notice, after which this version may no longer be accessible. 
\\ \\ 

\textbf{Copyright:}
\textcopyright~$2026$~IEEE. Personal use of this material is permitted. If the article gets accepted, permission from IEEE must be obtained for all other uses in any current or future media, including reprinting/republishing this material for advertising or promotional purposes, creating new collective works, for resale or redistribution to servers or lists, or reuse of any copyrighted component of this work in other works.

\end{tcolorbox}
\vspace{-5.8mm} 
\end{figure*}


\title{Feasibility-Aware Deep Reinforcement Learning for O-CU and O-DU Mapping in 6G O-RAN
\vspace{-0.40mm}}

\author{Mohammad Asif Habibi\orcidlink{0000-0001-9874-0047},~\IEEEmembership{Member,~IEEE}, 
Xavier Costa-P\'erez\orcidlink{0000-0002-9654-6109},~\IEEEmembership{Senior Member,~IEEE}, 
and Hans D. Schotten\orcidlink{0000-0001-5005-3635},~\IEEEmembership{Member,~IEEE}
\vspace{-11.8mm}
\thanks{\par\noindent\rule{0.99\columnwidth}{0.4pt}}
\thanks{This manuscript was received on DD MM $\mathrm{2027}$, revised on DD MM $\mathrm{2027}$, and accepted on DD MM $\mathrm{2027}$. It was published on DD MM $\mathrm{2027}$, and the current version was last updated on DD MM $\mathrm{2027}$. Mohammad Asif Habibi\orcidlink{0000-0001-9874-0047} and Hans D. Schotten\orcidlink{0000-0001-5005-3635} are with the Intelligent Networks (IN) Research Department, German Research Center for Artificial Intelligence (DFKI), $\mathrm{67663}$ Kaiserslautern, Germany. Xavier Costa-P\'erez\orcidlink{0000-0002-9654-6109} is with the Space and Terrestrial Systems Group, NEC Laboratories Europe, $\mathrm{69115}$ Heidelberg, Germany. He is also associated with the Department of AI-Driven Systems of the i$2$CAT Research Center and the Department of Engineering Sciences of the Catalan Institution for Research and Advanced Studies (ICREA), $\mathrm{08034}$ Barcelona, Spain. Hans D. Schotten\orcidlink{0000-0001-5005-3635} is also affiliated with the Division of Wireless Communications and Radio Navigation (WiCoN), Department of Electrical and Computer Engineering (EIT), University of Kaiserslautern (RPTU), $\mathrm{67663}$ Kaiserslautern, Germany. The corresponding author is Mohammad Asif Habibi\orcidlink{0000-0001-9874-0047} (\texttt{mohammad\_asif.habibi@dfki.de}).}
}

\maketitle
\begin{abstract}
This paper studies the mapping of the virtual network function components (VNFCs) of the O-RAN central unit (O-CU) and O-RAN distributed unit (O-DU) onto virtual machines (VMs) in sixth-generation (6G) open radio access networks (O-RAN). The problem is formulated as a feasibility-aware sequential decision process that maximizes the number of placed VNFCs and, secondarily, favors fewer active VMs and lower residual resource wastage. Although conventional optimization techniques can solve this problem, their computational cost becomes prohibitive for larger instances or when optimization must be repeated. To address this limitation, we explore two neural reinforcement learning (RL)-based methods: a deep $Q$-network (DQN) and a double deep $Q$-network (Double DQN). Whereas DQN uses the same target network to both select and evaluate the bootstrap action, Double DQN decouples these two operations to mitigate maximization-induced overestimation. We compare both methods against six baselines: random feasible placement, first-fit, first-fit decreasing, best-fit, a resource-wastage-aware heuristic, and a time-bounded mixed-integer linear programming (MILP) solver. We conduct a controlled experimental evaluation across four problem sizes with $30$ independent instances each, together with capacity-stress, distribution-shift, and reward-ablation experiments. For the $16 \times 8$ VNFC-to-VM scenario, DQN and Double DQN achieve mean placed-VNFC fractions of $0.990$ and $0.996$, respectively. However, first-fit decreasing and the resource-wastage-aware heuristic match this placement quality with lower efficiency cost and lower decision time, whereas the MILP attains the lowest efficiency cost at a much higher runtime. The results clarify the strengths and limitations of neural placement for this static problem, without establishing superiority over conventional methods.
\end{abstract}

\IEEEoverridecommandlockouts
\begin{keywords}
6G, deep $Q$-network, double deep $Q$-network, O-RAN, VNF placement, VNFC-to-VM mapping.
\vspace{-2.5mm}
\end{keywords}

\IEEEpeerreviewmaketitle

\section{Introduction}\label{Sec:Introduction}
\IEEEPARstart{T}{he} evolution toward \ac{6G} wireless networks is expected to enable highly dynamic, scalable, and intelligent communication systems that support heterogeneous service requirements, including ultra-low latency, massive connectivity, and ultra-high reliability \cite{9349624}. These stringent and often conflicting requirements motivate a shift from rigid, monolithic network architectures toward flexible, software-driven designs. In this context, the \ac{O-RAN} architecture has emerged as a key enabler of flexibility and programmability \cite{WG1ORANArchitecture}: it disaggregates traditional base stations into modular, interoperable components, namely the \ac{O-CU}, \ac{O-DU}, and \ac{O-RU}, which are connected through open interfaces \cite{11672720}. This functional decomposition promotes multi-vendor interoperability, can reduce deployment costs, and enables fine-grained control over \acp{NF} \cite{11664121}. Furthermore, when combined with virtualization and cloud-native principles, this disaggregation allows \acp{NF} to be implemented as \acp{VNF} and deployed on geographically distributed \ac{O-Cloud} infrastructure, thereby supporting dynamic resource allocation, network slicing, and on-demand service provisioning \cite{10090468}.

However, the efficient mapping of the \ac{O-CU} and \ac{O-DU}, as \acp{VNF}, onto available \acp{VM} in \ac{O-Cloud} sites remains a fundamental challenge \cite{9750106}. This problem is inherently complex because of the tight coupling among computational, networking, and radio resources \cite{10090468} and the need to satisfy diverse performance constraints such as latency, throughput, and reliability. The challenge is further exacerbated by \ac{O-RAN} slicing, where multiple coexisting slices impose heterogeneous and often conflicting requirements on a shared wireless infrastructure \cite{9750106}. Each slice may demand specific \ac{QoS} guarantees, isolation levels, and resource elasticity, turning the mapping problem into a multi-objective optimization task with stringent constraints. In such environments, suboptimal mapping decisions can lead to resource fragmentation, increased end-to-end latency, \ac{SLA} violations, and limited scalability of \ac{6G} networks \cite{10712647}. Underlying all of these formulations is a common core problem: deciding, under storage and compute capacity constraints, which \ac{VM} should host each \ac{VNF} component. This paper focuses on that core research problem.

\subsection{Prior Studies}\label{subsub:priorworks}
The \ac{VNF} mapping problem has been extensively investigated across various network environments, including wireless and cloud-native infrastructures, using a wide range of conventional optimization and learning-based methodologies \cite{10090468}. Because \ac{VNF} placement is NP-hard and subject to strict constraints on latency, resource utilization, and energy efficiency, classical heuristic and metaheuristic methods have been widely adopted to obtain good solutions at acceptable computational cost \cite{10090468}. For example, \ac{ACO} has been applied to joint \ac{VNF} placement and routing in \ac{SFC} deployment \cite{10229012}. In \ac{ACO}, artificial ants construct candidate mappings probabilistically, guided by pheromone trails that are reinforced by high-quality solutions. Mao \textit{et al.}~\cite{10229012} combine this mechanism with online learning to adapt deployment decisions over time while balancing objectives such as cost, service delay, and load balancing.

\Ac{MILP} is one of the common approaches to the \ac{VNF} mapping problem, since it can yield provably optimal solutions under resource and \ac{QoS} constraints. Representative works include Moens and De Turck~\cite{7014205}, who introduced an \ac{MILP}-based model for \ac{NFV} resource allocation; Bari \textit{et al.}~\cite{7367338}, who investigated \ac{ILP}-based and heuristic orchestration methods; and Mehraghdam \textit{et al.}~\cite{6968961}, who formalized the placement of \ac{VNF} chains as a mixed-integer quadratically constrained program. Martín-Pérez and Bernardos~\cite{8436765} extended the problem to multiple administrative domains, proposing scalable mapping algorithms that optimize inter-domain resource allocation under domain-level constraints and coordination overhead, which is particularly relevant for large-scale \ac{5G} and \ac{6G} networks spanning several operators. However, exact formulations become computationally intractable for large instances with complex constraints \cite{11157765}. Decomposition techniques such as column generation, which split the original problem into a master problem and pricing subproblems, improve tractability while preserving good solution quality \cite{10293056, 8314693}.

Recently, a growing body of research has turned to \ac{ML}-driven approaches, motivated by their potential to enable adaptive, data-driven resource management in cloud-native networks \cite{10090468}. In particular, \ac{RL} has attracted significant attention because \ac{VNF} placement is naturally a sequential decision-making problem under uncertainty: an agent learns placement policies through repeated interaction with the network environment, without requiring an explicit model of its dynamics \cite{8932445, 8873660}.

Building on this direction, our previous work \cite{Asif008Paper} introduced \texttt{SliceMapper}, an \ac{RL}-based framework that maps the functionalities of the \ac{O-CU} and \ac{O-DU}, referred to as \acp{VNFC}, onto \acp{VM}. It investigated four temporal-difference learning methods, namely the tabular and linear variants of \ac{SARSA} and $Q$-learning. The results demonstrated that \ac{RL} can solve the \ac{VNFC} mapping problem effectively, but they also exposed the limits of these representations: tabular methods require state-action values whose number grows with the problem size, whereas linear methods restrict the action-value function to a linear combination of hand-crafted features. Whether nonlinear function approximation, combined with experience replay and target networks, improves placement quality, efficiency, or learning stability in this setting therefore remained an open research question.

Other studies have applied deep \ac{RL} to \ac{VNF} mapping in \ac{SDN}/\ac{NFV} and \ac{5G} networks. Quang \textit{et al.}~\cite{8873660} embedded \ac{VNF} forwarding graphs with a deep \ac{RL} agent, and Pei \textit{et al.}~\cite{8932445} used deep \ac{RL} for \ac{VNF} placement in \ac{SDN}/\ac{NFV}-enabled networks. Kuai \textit{et al.}~\cite{9906071} applied proximal policy optimization to fair \ac{VNF} mapping and scheduling, Tran \textit{et al.}~\cite{10623413} proposed a deep \ac{RL} framework for \ac{5G} \ac{SFC} provisioning, and Wei \textit{et al.}~\cite{10946227} adapted \ac{SFC} mapping with multi-task deep \ac{RL}. These works confirm the growing use of deep \ac{RL} for resource allocation. However, none of them addresses the placement of the \ac{O-CU} and \ac{O-DU} \acp{VNFC} onto \acp{VM} in \acp{O-Cloud}. Moreover, deep \ac{RL} in general still faces challenges in training stability, computational overhead, and generalization across deployment scenarios.

\subsection{Major Research Problems} \label{Subsec:Problem}
The above studies, however, predominantly model \acp{VNF} as monolithic entities during the mapping process. Such an abstraction does not align with the fine-grained, cloud-native design principles of \ac{O-RAN}, for two reasons. First, each \ac{VNF} is composed of multiple \acp{VNFC} \cite{ETSI_NFV003}; in the \ac{O-CU} and \ac{O-DU}, these correspond to protocol-layer functions with distinct compute and storage requirements \cite{3GPP_TR38801}. A monolithic model must place all of them on a single host, whereas a component-level model admits every such co-located placement as well as placements that distribute the components across \acp{VM}. Component-level mapping can therefore only enlarge the set of feasible placements, enabling higher resource utilization and more flexible placement decisions, although distributed placements may incur additional inter-component communication overhead. Second, the resource demands of these components can vary considerably with service requirements and \ac{QoS} constraints, which a monolithic model cannot represent.

Furthermore, the literature typically treats the placement layers in isolation. Cloud-computing-related studies focus on \ac{VM}-to-\ac{PM} mapping and treat the \acp{VM} as given \cite{IaaSPaper}, whereas \ac{VNF} placement-related studies map entire \acp{VNF} onto physical servers or network nodes \cite{6968961, 8932445}. Neither perspective models the intermediate step of assigning \acp{VNFC} to \acp{VM}. When \acp{VM} are provisioned in advance without regard to the components they will host, the storage and compute capacity left unused within each \ac{VM}, i.e., its internal fragmentation, is not optimized. Explicitly modeling \ac{VNFC}-to-\ac{VM} mapping makes this residual capacity a decision variable, which we quantify through the active-\ac{VM} residual wastage in Section~\ref{Sec:PerformanceEvaluation}.

These observations motivate formulating \ac{VNFC}-to-\ac{VM} mapping explicitly as an optimization problem, which is the focus of this paper. Specifically, the mapping is many-to-one: each \ac{VNFC} of the \ac{O-CU} and \ac{O-DU}, characterized by its storage and compute demands, is either assigned to exactly one candidate \ac{VM} or rejected, whereas a \ac{VM} may host multiple \acp{VNFC} as long as their aggregate demands do not exceed its storage and compute capacities. The primary objective is to maximize the number of placed \acp{VNFC}; among placements of equal size, those that use fewer active \acp{VM} and leave less residual capacity are preferred. Because each assignment reduces the capacity available to subsequent components, the problem can be cast as a sequential decision process, which makes it amenable to \ac{RL}.

\subsection{Summary of Key Contributions}\label{Subsec:Contributions}
To address the aforementioned research problems, this paper extends our prior work \cite{Asif008Paper}, in which the \acp{VNFC} of the \ac{O-CU} and \ac{O-DU} were mapped onto \acp{VM} using tabular and linear variants of \ac{SARSA} and $Q$-learning. Building on the same system model, this study investigates two deep \ac{RL} methods, namely \ac{DQN} and Double \ac{DQN}. The main contributions are as follows:

\begin{itemize}
    \item We extend our prior work \cite{Asif008Paper} with \ac{DQN} and Double \ac{DQN} agents that share a $15$-dimensional action-conditioned feature representation, whose number of parameters is independent of the number of candidate \acp{VM}. Both agents combine feasible-action masking, multi-step bootstrap targets that account for forced rejections, experience replay, and target networks. 
    \item We provide a comparative benchmark of both agents, each with independently tuned hyperparameters, against random feasible placement, first-fit, first-fit decreasing, best-fit, a resource-wastage-aware heuristic, and a time-bounded \ac{MILP}. The benchmark covers four problem sizes with $30$ independent instances each and reports placement quality, resource efficiency, decision latency, training time, and memory usage.    
    \item We assess robustness and design choices through capacity-stress and zero-shot distribution-shift scenarios and a terminal-reward ablation. The results show that both agents achieve near-complete placement, but strong heuristics match their placement quality at lower efficiency cost and computational overhead. We therefore report both the favorable and the unfavorable outcomes of neural placement in the paper.
\end{itemize}

\subsection{The Structure of the Paper}\label{Subsec:Structure}
The rest of this paper is organized as follows. Section~\ref{Sec:SystemModel} presents the system model. Section~\ref{Sec:ProblemFormulation} formulates the \ac{VNFC}-to-\ac{VM} mapping problem, including its capacity constraints, the normalized efficiency cost, and the \ac{MILP} formulation. Section~\ref{Sec:ProposedSolution} motivates the use of \ac{RL}, casts the problem as a sequential decision process, and introduces the \ac{DQN} and Double \ac{DQN} agents together with their pseudocode. Section~\ref{Sec:PerformanceEvaluation} describes the experimental setup and compares both agents with heuristic and \ac{MILP} baselines in terms of placement quality, efficiency, and computational cost; it also examines failure conditions, distribution shifts, reward ablation, the comparison with our earlier tabular and linear methods, and the limitations of the evaluation. Finally, Section~\ref{Sec:Conclusion} concludes the paper and outlines directions for future research.


\section{System Model}\label{Sec:SystemModel}
\begin{figure}[ht]
    \centering
    \includegraphics[width=\columnwidth]{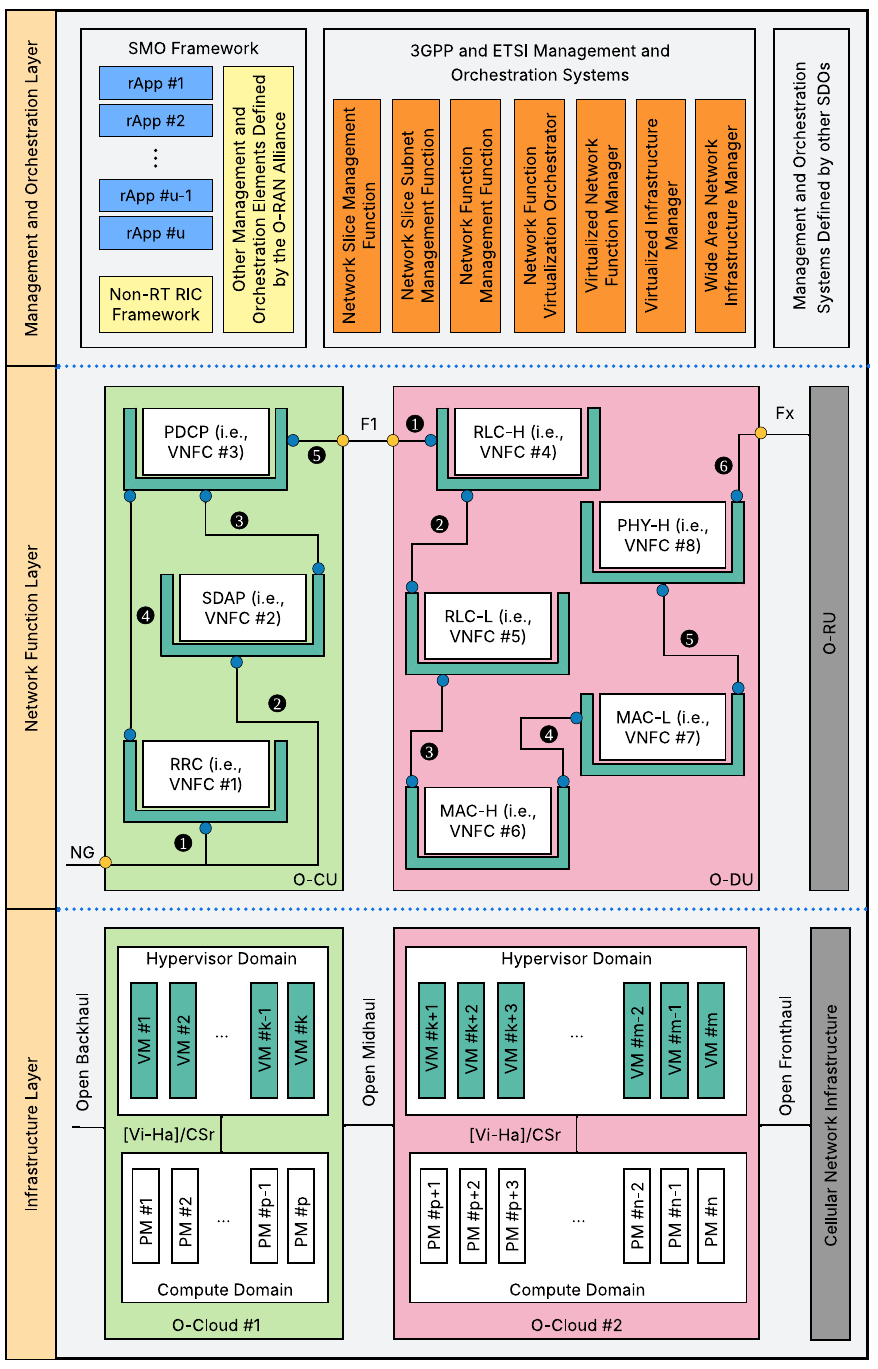}
    \caption{System model for mapping the \acp{VNFC} of the \ac{O-CU} and \ac{O-DU} onto \acp{O-Cloud}. The numbered lines denote the internal virtual links between \acp{VNFC}. \acp{VM}~\#1 to \#$k$ reside in \ac{O-Cloud}~\#1 and \acp{VM}~\#$k{+}1$ to \#$m$ in \ac{O-Cloud}~\#2; they form the pool $\mathcal{V}$ of $m$ candidate \acp{VM}. The optimization addresses the assignment of \acp{VNFC} to \acp{VM}; the eight \acp{VNFC} shown are illustrative, whereas we consider an arbitrary number $h$. RRC: radio resource control; SDAP: service data adaptation protocol; PDCP: packet data convergence protocol; RLC-H/RLC-L: radio link control-high/-low; MAC-H/MAC-L: medium access control-high/-low; PHY-H: physical-high.}
   \label{Fig:SystemModelDQL}
   \vspace{0.9mm}
\end{figure}

The system model is illustrated in Figure~$\mathrm{\ref{Fig:SystemModelDQL}}$. The architecture is organized into three logical layers: the Management and Orchestration Layer, the Network Function Layer, and the Infrastructure Layer. The Management and Orchestration Layer encompasses the \ac{SMO} framework defined by the \ac{O-RAN} Alliance, including the \ac{Non-RT RIC}, together with management components specified by other \acp{SDO}. These components manage and orchestrate the \acp{NF} of \ac{O-RAN} slices and their underlying resources within a non-real-time control loop that operates on timescales of 1~s or longer \cite{BuildingSMOFramework}. The Network Function Layer comprises the functional elements of an \ac{O-gNB}. Under the \ac{O-RAN} split~$7.2$x \cite{WG1ORANArchitecture}, the \ac{O-gNB} of an \ac{O-RAN} slice is decomposed into an \ac{O-CU}, an \ac{O-DU}, and an \ac{O-RU}, which are interconnected through the F1 interface and the open fronthaul (i.e., Fx), while the \ac{O-CU} connects to the core network through the NG interface. Finally, the Infrastructure Layer consists of the \ac{O-Cloud} sites, cell sites, and transport links. The \ac{O-Cloud} sites host the \ac{O-CU} and \ac{O-DU}, and the cell sites host the \ac{O-RU}. Each \ac{O-Cloud} site comprises a set of \acp{VM} hosted on \acp{PM}, and each \ac{VM} provides virtualized compute and storage resources to the \acp{VNFC} of the \ac{O-CU} and \ac{O-DU}. The \ac{O-RU} is outside the scope of this work. This paper considers a single \ac{O-RAN} slice subnet and focuses on the Infrastructure Layer: the Management and Orchestration Layer is assumed to execute the resulting placement decisions, and the Network Function Layer defines which \acp{VNFC} must be placed.

To formally model the functional decomposition introduced above, let $\mathcal{F}=\mathcal{C}\cup\mathcal{D}=\{f_1,\ldots,f_h\}$ denote the set of $h$ \acp{VNFC}, indexed in the order in which they are processed, where $\mathcal{C}$ and $\mathcal{D}$ contain the \ac{O-CU} and \ac{O-DU} components, respectively, with $\mathcal{C}\cap\mathcal{D}=\emptyset$. Each $f_i\in\mathcal{F}$ is characterized by its storage demand $S_i^{\mathrm{req}}$ and compute demand $C_i^{\mathrm{req}}$. Accordingly, the aggregate storage and compute demands are
\begin{align}
\mathbb{T}_{\mathrm{sto}}^{\mathrm{req}}
&=
\sum_{f_i\in\mathcal{C}} S_i^{\mathrm{req}}
+
\sum_{f_i\in\mathcal{D}} S_i^{\mathrm{req}},
\label{eq:total_storage_010}\\
\mathbb{T}_{\mathrm{com}}^{\mathrm{req}}
&=
\sum_{f_i\in\mathcal{C}} C_i^{\mathrm{req}}
+
\sum_{f_i\in\mathcal{D}} C_i^{\mathrm{req}},
\label{eq:total_compute_010}
\end{align}
which determine the resource load of an instance in Section~\ref{Sec:PerformanceEvaluation}.

Although Figure~\ref{Fig:SystemModelDQL} depicts eight functional components, the placement formulation admits an arbitrary number $h$ of \acp{VNFC}, each of which is treated as a placement request. This generalization accommodates the different problem sizes considered in Section~\ref{Sec:PerformanceEvaluation}. In the synthetic instances, each request is labeled as an \ac{O-CU} or \ac{O-DU} component according to the adopted decomposition, without explicitly simulating the individual protocol-layer functions. These labels do not enter the capacity constraints; they are used only to construct the \ac{O-CU}- and \ac{O-DU}-heavy demand profiles in Section~\ref{Sec:PerformanceEvaluation}.

The Infrastructure Layer comprises the virtualized resources available for hosting the \acp{VNFC}. Let $\mathcal{V}=\{v_1,\ldots,v_m\}$ denote the pool of $m$ candidate \acp{VM}, hosted on the underlying \acp{PM} of the \ac{O-Cloud} sites, where \ac{VM} $v_j$ has storage capacity $S_j^{\mathrm{cap}}$ and compute capacity $C_j^{\mathrm{cap}}$. The \ac{VNFC}-to-\ac{VM} mapping is modeled as a many-to-one assignment problem: each successfully placed \ac{VNFC} is assigned to exactly one candidate \ac{VM}, whereas a \ac{VM} may accommodate multiple \acp{VNFC} provided that sufficient residual storage and compute capacities remain. Following \cite{ETSI_NFV003}, each \ac{VNFC} instance runs in its own virtualization container (e.g., an operating-system container), and several such containers can be co-hosted on a single \ac{VM}; this paper allocates storage and compute resources at the \ac{VM} level. The requests are processed sequentially, with request $f_{t+1}$ handled at placement step $t\in\{0,\ldots,h-1\}$. Let $r_{j,t}^{S}$ and $r_{j,t}^{C}$ denote the residual storage and compute capacities of $v_j$ at step $t$, respectively, initialized as $r_{j,0}^{S}=S_j^{\mathrm{cap}}$ and $r_{j,0}^{C}=C_j^{\mathrm{cap}}$. A placement of $f_{t+1}$ onto $v_j$ is feasible if
\begin{equation}
\begin{aligned}
S_{t+1}^{\mathrm{req}} &\leq r_{j,t}^{S},\\
C_{t+1}^{\mathrm{req}} &\leq r_{j,t}^{C}.
\end{aligned}
\label{eq:vm_feasibility_010}
\end{equation}

If $f_{t+1}$ is placed on the selected \ac{VM} $v_j$, its residual capacities are updated as
\begin{equation}
r_{j,t+1}^{S}=r_{j,t}^{S}-S_{t+1}^{\mathrm{req}},
\qquad
r_{j,t+1}^{C}=r_{j,t}^{C}-C_{t+1}^{\mathrm{req}},
\label{eq:residual_update_010}
\end{equation}
while the residual capacities of all other \acp{VM} remain unchanged. If no \ac{VM} satisfies \eqref{eq:vm_feasibility_010}, the request is rejected and all residual capacities remain unchanged.

The objectives and constraints of this mapping problem are formulated in Section~\ref{Sec:ProblemFormulation}, and the sequential decision process is defined in detail in Section~\ref{Subsec:RLSolution}.


\section{Problem Formulation} \label{Sec:ProblemFormulation}
This section formulates the \ac{VNFC}-to-\ac{VM} mapping optimization problem introduced in Section~\ref{Sec:SystemModel}, following the formulation in \cite{Asif008Paper}. The $h$ \acp{VNFC} with storage and compute demands $S_i^{\mathrm{req}}$ and $C_i^{\mathrm{req}}$ must be assigned to the $m$ candidate \acp{VM} with capacities $S_j^{\mathrm{cap}}$ and $C_j^{\mathrm{cap}}$, such that the aggregate demands hosted on each \ac{VM} do not exceed either capacity. The optimization problem is therefore a two-dimensional vector bin-packing problem in which the number of placed items is maximized \cite{BinPackingRef}. Since each placement reduces the residual capacities of the selected \ac{VM}, the feasibility of every subsequent placement depends on the earlier decisions.

For each $f_i\in\mathcal{F}$ and $v_j\in\mathcal{V}$, let the binary variables $x_{ij}$, $y_j$, and $z_i\in\{0,1\}$ indicate whether $f_i$ is mapped to $v_j$, whether $v_j$ is active, and whether $f_i$ is placed, respectively. The feasibility constraints are
\begin{align}
 \sum_{j=1}^{m}x_{ij} &= z_i,
 && i=1,\ldots,h,
 \label{eq:assignment-revised}\\
 \sum_{i=1}^{h}S_i^{\mathrm{req}}x_{ij}
     &\le S_j^{\mathrm{cap}}y_j,
 && j=1,\ldots,m,
 \label{eq:storage-revised}\\
 \sum_{i=1}^{h}C_i^{\mathrm{req}}x_{ij}
     &\le C_j^{\mathrm{cap}}y_j,
 && j=1,\ldots,m,
 \label{eq:compute-revised}\\
 x_{ij} &\le y_j,
 && i=1,\ldots,h,\;j=1,\ldots,m,
 \label{eq:vm-active-link-1}\\
 y_j &\le \sum_{i=1}^{h}x_{ij},
 && j=1,\ldots,m.
 \label{eq:vm-active-link-2}
\end{align}
Constraint~\eqref{eq:assignment-revised} assigns each placed \ac{VNFC} to exactly one \ac{VM}, and \eqref{eq:storage-revised}--\eqref{eq:compute-revised} enforce the storage and compute capacities of active \acp{VM}. Constraints~\eqref{eq:vm-active-link-1}--\eqref{eq:vm-active-link-2} ensure that $y_j=1$ exactly when $v_j$ hosts at least one \ac{VNFC}; \eqref{eq:vm-active-link-1} is implied by \eqref{eq:storage-revised} for positive demands tightens the relaxation.

A mapping is complete if $z_i=1$ for all $i\in\{1,\ldots,h\}$. For a given mapping, the residual storage and compute capacities of $v_j$ are
\begin{align}
r_j^S = r_{j,h}^S
&= S_j^{\mathrm{cap}}
   - \sum_{i=1}^{h} S_i^{\mathrm{req}} x_{ij},
\label{eq:residual_storage_final}\\
r_j^C = r_{j,h}^C
&= C_j^{\mathrm{cap}}
   - \sum_{i=1}^{h} C_i^{\mathrm{req}} x_{ij},
\label{eq:residual_compute_final}
\end{align}
which coincide with the residual capacities obtained after all $h$ requests have been processed by the sequential updates in \eqref{eq:residual_update_010}.

Let $U=\sum_{j=1}^{m} y_j$ denote the number of active \acp{VM}, and let $M=\min\{h,m\}$. We define the aggregate residual fraction of the active \acp{VM} as
\begin{equation}
W=\frac{1}{2}\sum_{j=1}^{m} y_j\left(\frac{r_j^S}{S_j^{\mathrm{cap}}}+\frac{r_j^C}{C_j^{\mathrm{cap}}}\right),
\label{eq:normalized-waste}
\end{equation}
where normalizing each residual by the corresponding capacity makes storage and compute, which are measured in different units, directly comparable. The normalized efficiency cost is
\begin{equation}
 J_{\beta}=\frac{(1-\beta)U+\beta W}{M},
 \qquad 0\leq\beta\leq1.
 \label{eq:efficiency-cost}
\end{equation}
Since each active \ac{VM} hosts at least one \ac{VNFC}, $U\le M$, and since each residual fraction lies in $[0,1]$, $W\le U$; hence $0\le J_\beta\le 1$. Among placements with the same number of placed \acp{VNFC}, a smaller $J_\beta$ favors fewer active \acp{VM} and lower residual capacity. In all experiments, $\beta=0.5$. To report resource wastage, we additionally use the mean residual fraction over active \acp{VM},
\begin{equation}
\label{eq:W2}
    W_{\mathrm{act}} = \begin{cases}
        W/U, & \text{if } U>0,\\[1.5mm]
        1,   & \text{if } U = 0,
    \end{cases}
\end{equation}
where the value $1$ for $U=0$ is a convention assigning the worst wastage when no \ac{VNFC} is placed.

The \ac{MILP} minimizes
\begin{equation}
\label{eq:milp-objective}
\min_{x,y,z}
\quad
J_\beta
-
\kappa\sum_{i=1}^{h}z_i,
\end{equation}
subject to \eqref{eq:assignment-revised}--\eqref{eq:vm-active-link-2}, where $\kappa = 2$ is the dominance coefficient. Since $0\le J_\beta\le 1$, placing one additional \ac{VNFC} decreases the second term of \eqref{eq:milp-objective} by $\kappa$, whereas $J_\beta$ can increase by at most $1$. Hence, any solution that places more \acp{VNFC} attains a strictly lower objective value, and any $\kappa>1$ would suffice. The \ac{MILP} maximizes the number of placed \acp{VNFC} and, among solutions with the same placement count, minimizes the efficiency cost. Substituting \eqref{eq:residual_storage_final}--\eqref{eq:efficiency-cost} and using $x_{ij}y_j=x_{ij}$, which follows from \eqref{eq:vm-active-link-1}, yields the equivalent linear objective
\begin{align}
\min_{x,y,z}\quad
&
\frac{1}{M}\sum_{j=1}^{m}y_j
-
\frac{\beta}{2M}
\sum_{i=1}^{h}\sum_{j=1}^{m}
x_{ij}
\left(
\frac{S_i^{\mathrm{req}}}{S_j^{\mathrm{cap}}}
+
\frac{C_i^{\mathrm{req}}}{C_j^{\mathrm{cap}}}
\right)
\nonumber\\
&
-
\kappa\sum_{i=1}^{h}z_i.
\label{eq:milp-objective-expanded}
\end{align}

The \ac{MILP} is solved with SciPy's \texttt{milp} interface to the HiGHS solver, using a $30$~s time limit and a relative optimality-gap tolerance of $10^{-6}$. The solver status and reported gap are recorded to distinguish certified optimal solutions from incumbents returned at the time limit.

\begin{table}[t]
\caption{Notation used in the system model and the proposed solution}
\label{Tab:RLNotations}
\centering
\small
\renewcommand{\arraystretch}{1.40}
\begin{tabularx}{\columnwidth}{|l|X|}
\hline
\rowcolor{tableheader}
\multicolumn{1}{|c|}{\textbf{\textcolor{white}{Notation}}} & \multicolumn{1}{c|}{\textbf{\textcolor{white}{Description}}} \\ \hline

\multicolumn{2}{|l|}{\textit{System Model and Problem Formulation}} \\ \hline

$\mathcal{F}$, $f_i$ & Set of $h$ \acp{VNFC}; $i$-th \ac{VNFC} \\ \hline

$\mathcal{C}$, $\mathcal{D}$ & Sets of \ac{O-CU} and \ac{O-DU} \acp{VNFC} \\ \hline

$\mathcal{V}$, $v_j$ & Set of $m$ candidate \acp{VM}; $j$-th \ac{VM} \\ \hline

$S_i^{\mathrm{req}}$, $C_i^{\mathrm{req}}$ & Storage and compute demands of $f_i$ \\ \hline

$\mathbb{T}_{\mathrm{sto}}^{\mathrm{req}}$, $\mathbb{T}_{\mathrm{com}}^{\mathrm{req}}$ & Aggregate storage and compute demands \\ \hline

$S_j^{\mathrm{cap}}$, $C_j^{\mathrm{cap}}$ & Storage and compute capacities of $v_j$ \\ \hline

$t$ & Placement step, $t\in\{0,\ldots,h-1\}$ \\ \hline

$r_{j,t}^{S}$, $r_{j,t}^{C}$ & Residual capacities of $v_j$ at step $t$ \\ \hline

$r_{j}^{S}$, $r_{j}^{C}$ & Residual capacities of $v_j$ after all requests \\ \hline

$x_{ij}$ & Binary \ac{VNFC}-to-\ac{VM} assignment variable \\ \hline

$y_j$ & \ac{VM} activation indicator \\ \hline

$z_i$ & \ac{VNFC} placement indicator \\ \hline

$U$, $M$ & Number of active \acp{VM}; $\min\{h,m\}$ \\ \hline

$W$ & Aggregate residual fraction of active \acp{VM} \\ \hline

$W_{\mathrm{act}}$ & Mean residual fraction over active \acp{VM} \\ \hline

$J_\beta$, $\beta$ & Normalized efficiency cost; its weighting factor \\ \hline

$\kappa$ & Dominance coefficient of the \ac{MILP} objective \\ \hline

\multicolumn{2}{|l|}{\textit{Reinforcement Learning Formulation}} \\ \hline

$\mathbf{d}_t$, $\mathbf{D}_t$ & Current and remaining aggregate demand vectors \\ \hline

$S_t$, $A_t$ & Placement state and selected \ac{VM} at step $t$ \\ \hline

$\mathcal{A}$, $\mathcal{A}_f(S_t)$ & Action space; feasible actions in state $S_t$ \\ \hline

$R_{t+1}$ & Reward after processing request $f_{t+1}$ \\ \hline

$\mathbf{1}_t$, $\eta$ & Completion indicator; terminal efficiency-bonus coefficient \\ \hline

$k_t$, $\Gamma_t$ & Consecutive forced rejections; bootstrap discount \\ \hline

$\rho$ & Placed-\ac{VNFC} fraction \\ \hline

$E$ & Greedy-policy evaluation score \\ \hline

$\boldsymbol{\phi}(s,a)$, $p_t$ & $15$-dimensional state-action features; placement progress $t/h$ \\ \hline

$Q_\theta$, $Q_{\bar\theta}$ & Online and target action-value networks \\ \hline

$\theta$, $\bar\theta$ & Online and target network parameters \\ \hline

$\hat{\pi}$ & Greedy placement policy \\ \hline

$\gamma$ & Discount factor \\ \hline

$\varepsilon_e$ & Exploration rate in training episode $e$ \\ \hline

$\alpha$ & Learning rate \\ \hline

$B$, $N$, $N_w$ & Minibatch size, replay capacity, warm-up size \\ \hline

$F$, $K$ & Training interval; target-update interval \\ \hline

$E_{\max}$ & Maximum number of training episodes \\ \hline

\end{tabularx}
\end{table}


\section{Proposed Solution}\label{Sec:ProposedSolution}
Building on the formulation in Section~\ref{Sec:ProblemFormulation}, this section introduces two \ac{RL}-based solutions for \ac{VNFC}-to-\ac{VM} mapping: Double \ac{DQN} and \ac{DQN}. We first motivate the use of \ac{RL} and cast the mapping as a sequential decision process, and then present the two agents together with their training procedures. The key notation is summarized in Table~\ref{Tab:RLNotations}.

\subsection{Why Reinforcement Learning for VNFC Mapping?}\label{Subsec:WhyRL}
The \ac{MILP} in \eqref{eq:milp-objective} can be solved to optimality for small, static instances. Its cost, however, grows quickly with problem size: in our experiments, the solver certifies optimality within $30$~s for all $16\times8$ instances, but for only one of thirty $32\times16$ instances and for none at $64\times32$ (see Section~\ref{Sec:PerformanceEvaluation}). Practical \ac{O-RAN} deployments are, moreover, larger and dynamic: \ac{VNFC} demands vary over time, \ac{VM} capacities change with background workloads, and slice-level latency and bandwidth requirements differ across services, so the problem would have to be solved repeatedly. This paper addresses the static problem as a first step, in order to establish whether learned placement policies can match exact and heuristic methods before extending them to dynamic settings.

\ac{RL} is applicable because the \ac{VNFC}-to-\ac{VM} mapping can be decomposed into a sequence of decisions: at each step, the agent selects a \ac{VM} for the current \ac{VNFC}, and this choice changes the residual resources available to all subsequent requests. By maximizing a cumulative reward, an \ac{RL} agent can learn placement policies that account for these downstream effects rather than only immediate feasibility. Once trained, a policy produces each decision with one forward pass of its network per candidate \ac{VM}, which is far cheaper than solving the \ac{MILP}, although heuristics remain faster still. This inference advantage, however, materializes only when a trained policy is reused across instances; in our main benchmark, a separate policy is trained for each instance, and its training cost must therefore be weighed against a single \ac{MILP} solution. The role of \ac{RL} in this paper is thus to learn sequential, feasibility-aware placement policies and to quantify their placement quality, efficiency, and computational cost relative to exact and heuristic methods, including their behavior on unseen instances (refer to Section~\ref{Sec:PerformanceEvaluation} for more details).

The \ac{DQN} and Double \ac{DQN} agents build on our previous work \cite{Asif008Paper}, which applied tabular and linear variants of SARSA and $Q$-learning to the same problem. They extend these methods with nonlinear, neural-network-based action-value approximation, experience replay, and target networks.

\subsection{Sequential Decision Formulation}\label{Subsec:RLSolution}
The \ac{RL} agent processes the \acp{VNFC} sequentially at the placement steps $t\in\{0,\ldots,h-1\}$ defined in Section~\ref{Sec:SystemModel}. An episode is one complete pass over the entire request batch, and the complete ordered batch is known at the beginning of each episode. At step $t$, the demand vector of the current request is
\begin{equation}
\mathbf{d}_t
=
\left(S_{t+1}^{\mathrm{req}},C_{t+1}^{\mathrm{req}}\right),
\label{eq:demand-code-index}
\end{equation}
and the environment state is
\begin{equation}
S_t
=
\left(
t,\mathbf{d}_t,
\left\{
S_j^{\mathrm{cap}},C_j^{\mathrm{cap}},
r_{j,t}^{S},r_{j,t}^{C}
\right\}_{j=1}^{m}, \mathbf{D}_t
\right),
\label{eq:formal-state}
\end{equation}
which records the current request, the capacities of the candidate \acp{VM}, their residual capacities after preceding placements, and the aggregate demands of the unprocessed requests, including the current one,
\begin{equation}
\mathbf{D}_t=
\left(
\sum_{i=t+1}^{h}S_i^{\mathrm{req}},
\sum_{i=t+1}^{h}C_i^{\mathrm{req}}
\right).
\label{eq:remaining_demand010}
\end{equation}

At each step, an action selects one \ac{VM} for the current request, so the action space $\mathcal{A}=\{1,\ldots,m\}$ contains the indices of all candidate \acp{VM}. The agent is, however, restricted to the feasible subset
\begin{equation}
\mathcal A_f(S_t)
=
\left\{
j\in\mathcal{A} \mid
S_{t+1}^{\mathrm{req}}\leq r_{j,t}^{S},
\;
C_{t+1}^{\mathrm{req}}\leq r_{j,t}^{C}
\right\},
\label{eq:feasible-action-set}
\end{equation}
i.e., infeasible \acp{VM} are masked and never selected during training or evaluation. If $A_t=j\in\mathcal A_f(S_t)$ is selected, the residual capacities are updated according to \eqref{eq:residual_update_010}.

If $\mathcal A_f(S_t)=\varnothing$, the current request cannot be placed and is rejected; we refer to this as a forced rejection. Following \cite{Asif008Paper}, the reward received after step $t$ is
\begin{equation}
R_{t+1}
=
\begin{cases}
-1,
&
\mathcal A_f(S_t)=\varnothing,
\\[1.5mm]
\dfrac{1}{h}
+
\eta\,\mathbf{1}_t\,(1-J_\beta),
&
A_t\in\mathcal A_f(S_t),
\end{cases}
\label{eq:revised-reward}
\end{equation}
where $\mathbf{1}_t\in\{0,1\}$ equals one only if $t=h-1$ and the mapping is complete, $J_\beta$ is the efficiency cost \eqref{eq:efficiency-cost} of the final mapping, and $\eta=0.25$ is the terminal efficiency-bonus coefficient. The placement rewards of a complete mapping sum to one, so the bonus $\eta(1-J_\beta)\le\eta$ is granted only when all $h$ \acp{VNFC} are placed and favors more efficient mappings.

The placed-\ac{VNFC} fraction of a mapping is
\begin{equation}
\rho=
\frac{1}{h}\sum_{i=1}^{h}z_i,
\label{eq:placed_fraction010}
\end{equation}
with $z_i$ as defined in Section~\ref{Sec:ProblemFormulation}, and the greedy-policy evaluation score is
\begin{equation}
E=
\rho
+\eta\,
\mathbf{1}_{\{\rho=1\}}
(1-J_\beta),
\label{eq:evaluation_score010}
\end{equation}
where $\rho=1$ corresponds to a complete mapping. Hence, $E\in[0,1+\eta]$: it exceeds one only for complete mappings, among which it prefers smaller efficiency costs. The score is used for checkpoint evaluation and for the stabilization criterion in Section~\ref{Sec:PerformanceEvaluation}; it is distinct from the discounted temporal-difference return and from the \ac{MILP} objective \eqref{eq:milp-objective}.

The \ac{DQN} and Double \ac{DQN} agents share the placement environment, the feasible-action constraints, and the reward formulation; they differ only in their bootstrap targets. Both use an action-conditioned representation: for every candidate $a\in\mathcal A$, a feature vector $\boldsymbol{\phi}(S_t,a)\in\mathbb R^{15}$ is constructed, and the corresponding action value is approximated by
\begin{equation}
Q_{\theta}(S_t,a)=
f_{\theta}
\bigl(\boldsymbol{\phi}(S_t,a)\bigr),
\label{eq:qnetwork010}
\end{equation}
where $f_{\theta}$ is a fully connected network with online parameters $\theta$, two hidden layers of $128$ ReLU units, and a scalar output, i.e., $18{,}689$ trainable parameters. Because the network scores one state-action pair at a time, its number of parameters is independent of the number of candidate \acp{VM}~$m$. During evaluation, the agent follows the greedy policy
\begin{equation}
\hat{\pi}(S_t)=
\arg\max_{a\in\mathcal A_f(S_t)}
Q_{\theta}(S_t,a),
\label{eq:greedy_action010}
\end{equation}
which selects the feasible \ac{VM} with the highest action value.

As in \cite{Asif008Paper}, $\boldsymbol{\phi}(S_t,a)$ comprises $15$ features in three groups. The eight candidate-\ac{VM} features are the residual storage and compute capacities of \ac{VM}~$a$ before and after a hypothetical placement of the current request, a feasibility flag, the activation status of \ac{VM}~$a$, the post-placement fit $1-\tfrac{1}{2}(\hat s_a+\hat c_a)$, and the storage-compute balance $1-|\hat s_a-\hat c_a|$, where $\hat s_a$ and $\hat c_a$ denote the post-placement residual storage and compute fractions of \ac{VM}~$a$. The four demand features are the normalized storage and compute demands of the current request and the remaining storage and compute demand pressures from $\mathbf{D}_t$. The three global features are the fraction of active \acp{VM}, the placement progress $p_t=t/h$, and a bias equal to one.

\subsection{Double Deep $Q$-Network}\label{Subsec:DDQN}
Double \ac{DQN} is an off-policy deep \ac{RL} method. Like Double $Q$-learning, it separates action selection from action evaluation when constructing bootstrap targets, which reduces the overestimation caused by maximization \cite{vanHasselt2016DoubleDQN, 29970462997187}. Both agents employ an online network $Q_{\theta}$ and a target network $Q_{\bar{\theta}}$, where the target network is initialized with $\bar{\theta}=\theta$ and periodically synchronized during training. The online network is trained by gradient descent, and the target network provides the bootstrap values. In Double \ac{DQN}, the online network selects the next action and the target network evaluates it.

States with $\mathcal A_f(S_t)=\varnothing$ require no decision. The agents learn only from decision states, and the forced rejections between two decision states are folded into a single transition. Let $k_t$ denote the number of consecutive forced rejections following the action at step $t$. The accumulated discounted reward and the bootstrap discount of this transition are
\begin{equation}
R_t^{(k_t)}=
\sum_{\ell=0}^{k_t}
\gamma^\ell R_{t+\ell+1},
\qquad
\Gamma_t=\gamma^{k_t+1},
\label{eq:accumulated_reward}
\end{equation}
so the negative rewards of forced rejections are included, and the discount accounts for the steps until the next decision state. For $k_t=0$, these reduce to the one-step reward and discount.

Each stored transition $b$ comprises the decision state $S_b$, the action $A_b$, the accumulated reward $R_b^{(k_b)}$, the discount $\Gamma_b$, and the next decision state $S'_b$. In Double \ac{DQN}, the online network selects the next action,
\begin{equation}
A_b^*=
\arg\max_{a'\in\mathcal A_f(S'_b)}
Q_{\theta}(S'_b,a'),
\label{eq:ddqn_action_selection010}
\end{equation}
and the target network evaluates it:
\begin{equation}
Y_b^{\mathrm{DDQN}}=
\begin{cases}
R_b^{(k_b)},
& \text{if } S'_b \text{ is terminal},\\[1mm]
R_b^{(k_b)}
+\Gamma_b\,Q_{\bar{\theta}}(S'_b,A_b^*),
& \text{otherwise}.
\end{cases}
\label{eq:ddqn_target010}
\end{equation}
This separation reduces overestimation, but does not guarantee unbiased value estimates or convergence to the optimal policy.

Both agents are trained with experience replay: transitions are stored in a replay buffer $\mathcal B_{\mathrm{rep}}$ of capacity $N$, from which minibatches of size $B$ are sampled uniformly. The online parameters minimize the Huber loss
\begin{equation}
\mathcal L(\theta)=
\frac{1}{B}
\sum_{b=1}^{B}
\operatorname{Huber}_{\delta}
\left(
Q_{\theta}(S_b,A_b)-Y_b
\right),
\label{eq:neural_huber010}
\end{equation}
where $Y_b$ is the agent-specific target, treated as a constant, and $\delta=1$ is the Huber threshold. The parameters are updated with the AdamW optimizer \cite{Loshchilov2019AdamW} and gradient clipping, and the target network is synchronized periodically. \texttt{Algorithm~\ref{alg:slicemapper-double-dqn}} summarizes the Double \ac{DQN} training procedure, where $\Phi'$ stacks the feature vectors $\boldsymbol{\phi}(S'_b,a)$ of all $a\in\mathcal A$, and $\mathcal M'$ is the binary mask of $\mathcal A_f(S'_b)$. The stopping criterion, hyperparameters, and training results are given in Section~\ref{Sec:PerformanceEvaluation}.

\begin{algorithm}[t]
\caption{Double \ac{DQN} training for \ac{VNFC}-to-\ac{VM} mapping}
\label{alg:slicemapper-double-dqn}
\begin{algorithmic}[1]
\footnotesize
\Require
\begin{itemize}
    \item Ordered \ac{VNFC} batch and \ac{VM} capacities
    \item Discount factor $\gamma$, exploration schedule $\varepsilon_e$
    \item Learning rate $\alpha$, minibatch size $B$
    \item Replay capacity $N$, warm-up threshold $N_w$
    \item Training interval $F$, target-update interval $K$
    \item Maximum number of training episodes $E_{\max}$
\end{itemize}
\State Initialize online network $Q_\theta$ randomly 
\State Initialize target network $Q_{\bar{\theta}}$ with $\bar{\theta}=\theta$
\State Initialize replay buffer $\mathcal{B}_{\mathrm{rep}}=\emptyset$ and AdamW optimizer
\State Initialize counters $n_{\mathrm{env}}=0$ and $n_{\mathrm{grad}}=0$
\For{$e=1,\ldots,E_{\max}$}
    \State Reset environment and process initial forced rejections
    \While{a nonterminal decision state $S_t$ exists}
        \State Determine $\mathcal{A}_f(S_t)$ and construct $\boldsymbol{\phi}(S_t,a)$ for all $a\in\mathcal{A}_f(S_t)$
        \State Select $\varepsilon_e$-greedy feasible action $A_t$ using $Q_\theta$
        \State Execute $A_t$ and process $k_t$ subsequent forced rejections
        \State Compute $R_t^{(k_t)}$ and $\Gamma_t$ using \eqref{eq:accumulated_reward}
        \State Store $(\boldsymbol{\phi}(S_t,A_t),R_t^{(k_t)},\Gamma_t,\Phi',\mathcal{M}',\mathrm{done})$ in $\mathcal{B}_{\mathrm{rep}}$
        \State $n_{\mathrm{env}}\leftarrow n_{\mathrm{env}}+k_t+1$
        \If{$|\mathcal{B}_{\mathrm{rep}}|\geq\max(B,N_w)$ \textbf{and} $n_{\mathrm{env}}\bmod F=0$}
            \State Sample $B$ transitions uniformly from $\mathcal{B}_{\mathrm{rep}}$
            \For{each sampled transition $b$}
                \If{$S'_b$ is terminal}
                    \State $Y_b^{\mathrm{DDQN}}=R_b^{(k_b)}$
                \Else
                    \State Select $A_b^*$ with $Q_\theta$ using \eqref{eq:ddqn_action_selection010}
                    \State Compute $Y_b^{\mathrm{DDQN}}$ with $Q_{\bar{\theta}}$ using \eqref{eq:ddqn_target010}
                \EndIf
            \EndFor
            \State Update $\theta$ by minimizing \eqref{eq:neural_huber010} using AdamW with gradient clipping
            \State $n_{\mathrm{grad}}\leftarrow n_{\mathrm{grad}}+1$
            \If{$n_{\mathrm{grad}}\bmod K=0$}
                \State $\bar{\theta}\leftarrow\theta$
            \EndIf
        \EndIf
    \EndWhile
    \State Evaluate greedy policy at scheduled checkpoints
    \If{stabilization criterion is satisfied after the minimum episode budget}
        \State \textbf{break}
    \EndIf
\EndFor
\State \Return trained online network $Q_\theta$ and greedy policy $\hat{\pi}$
\end{algorithmic}
\end{algorithm}

\subsection{Deep \texorpdfstring{$Q$}{Q}-Network}\label{Subsec:DQN}
\ac{DQN} is an off-policy deep \ac{RL} method that learns from \ac{TD} targets using experience replay and a target network \cite{Mnih2015DQN,Zhao2025}. Our \ac{DQN} agent shares all components described in Section~\ref{Subsec:DDQN}, i.e., the feature representation, network architecture, feasible-action masking, forced-rejection transitions, replay mechanism, and Huber loss \eqref{eq:neural_huber010}, and differs only in its target. Instead of decoupling selection and evaluation, \ac{DQN} uses the target network for both:
\begin{equation}
Y_b^{\mathrm{DQN}}=
\begin{cases}
R_b^{(k_b)},
& \text{if } S'_b \text{ is terminal},\\[1mm]
R_b^{(k_b)}
+\Gamma_b\displaystyle\max_{a'\in\mathcal A_f(S'_b)}Q_{\bar{\theta}}(S'_b,a'),
& \text{otherwise}.
\end{cases}
\label{eq:dqn_target}
\end{equation}
\texttt{Algorithm~\ref{alg:dqn}} summarizes the training procedure. Because the hyperparameters of both agents are tuned independently (Section~\ref{Sec:PerformanceEvaluation}), their empirical differences cannot be attributed exclusively to the target construction.

\begin{algorithm}[t]
\caption{\ac{DQN} training for \ac{VNFC}-to-\ac{VM} mapping}
\label{alg:dqn}
\begin{algorithmic}[1]
\footnotesize
\Require
\begin{itemize}
    \item Ordered \ac{VNFC} batch and \ac{VM} capacities
    \item Discount factor $\gamma$, exploration schedule $\varepsilon_e$
    \item Learning rate $\alpha$, minibatch size $B$
    \item Replay capacity $N$, warm-up threshold $N_w$
    \item Training interval $F$, target-update interval $K$
    \item Maximum number of training episodes $E_{\max}$
\end{itemize}
\State Initialize online network $Q_\theta$ randomly 
\State Initialize target network $Q_{\bar{\theta}}$ with $\bar{\theta}=\theta$
\State Initialize replay buffer $\mathcal{B}_{\mathrm{rep}}=\emptyset$ and AdamW optimizer
\State Initialize counters $n_{\mathrm{env}}=0$ and $n_{\mathrm{grad}}=0$
\For{$e=1,\ldots,E_{\max}$}
    \State Reset environment and process initial forced rejections
    \While{a nonterminal decision state $S_t$ exists}
        \State Determine $\mathcal{A}_f(S_t)$ and construct $\boldsymbol{\phi}(S_t,a)$ for all $a\in\mathcal{A}_f(S_t)$
        \State Select $\varepsilon_e$-greedy feasible action $A_t$ using $Q_\theta$
        \State Execute $A_t$ and process $k_t$ subsequent forced rejections
        \State Compute $R_t^{(k_t)}$ and $\Gamma_t$ using \eqref{eq:accumulated_reward}
        \State Store $(\boldsymbol{\phi}(S_t,A_t),R_t^{(k_t)},\Gamma_t,\Phi',\mathcal{M}',\mathrm{done})$ in $\mathcal{B}_{\mathrm{rep}}$
        \State $n_{\mathrm{env}}\leftarrow n_{\mathrm{env}}+k_t+1$
        \If{$|\mathcal{B}_{\mathrm{rep}}|\geq\max(B,N_w)$ \textbf{and} $n_{\mathrm{env}}\bmod F=0$}
            \State Sample $B$ transitions uniformly from $\mathcal{B}_{\mathrm{rep}}$
            \For{each sampled transition $b$}
                \State Compute $Y_b^{\mathrm{DQN}}$ with $Q_{\bar{\theta}}$ using \eqref{eq:dqn_target}
            \EndFor
            \State Update $\theta$ by minimizing \eqref{eq:neural_huber010} using AdamW with gradient clipping
            \State $n_{\mathrm{grad}}\leftarrow n_{\mathrm{grad}}+1$
            \If{$n_{\mathrm{grad}}\bmod K=0$}
                \State $\bar{\theta}\leftarrow\theta$
            \EndIf
        \EndIf
    \EndWhile
    \State Evaluate greedy policy at scheduled checkpoints
    \If{stabilization criterion is satisfied after the minimum episode budget}
        \State \textbf{break}
    \EndIf
\EndFor
\State \Return trained online network $Q_\theta$ and greedy policy $\hat{\pi}$
\end{algorithmic}
\end{algorithm}


\section{Performance Evaluation}\label{Sec:PerformanceEvaluation}
In this section, the two agents are compared with five conventional heuristics and a time-bounded \ac{MILP} baseline, following the controlled experimental framework of our previous work~\cite{Asif008Paper}. We examine placement quality, resource efficiency, empirical learning stabilization, online decision latency, and training cost. We further investigate the learned policies under capacity stress and distribution shifts, conduct a reward ablation, compare the agents with the tabular and linear methods of~\cite{Asif008Paper}, and discuss the limitations of the evaluation.

\subsection{Experimental Setup and Evaluation Metrics}\label{Subsec:SimulationsSettings}
The experiments are implemented in Python using NumPy, SciPy, and PyTorch. Four controlled problem sizes are considered, $(h,m)\in\{(8,4),(16,8),(32,16),(64,32)\}$. The \ac{VM} storage and compute capacities are sampled independently from $\mathcal{DU}\{80,\ldots,180\}$ and $\mathcal{DU}\{8,\ldots,20\}$, respectively. The \ac{VNFC} storage and compute demands are constructed from weights drawn independently from $\mathcal{U}(0.5,1.5)$ and scaled separately for each resource to a target load $\lambda_{\mathrm{sto}}=\mathbb{T}_{\mathrm{sto}}^{\mathrm{req}}/\sum_{j}S_j^{\mathrm{cap}}$ and $\lambda_{\mathrm{com}}=\mathbb{T}_{\mathrm{com}}^{\mathrm{req}}/\sum_{j}C_j^{\mathrm{cap}}$ of $0.75$. Each demand is capped at $90\%$ of the largest corresponding \ac{VM} capacity, so that every \ac{VNFC} fits into at least one empty \ac{VM}, and rounded to a positive integer; the realized loads may deviate slightly from the target. All capacities and demands are in synthetic units.

\begin{table}[ht!]
\caption{Experimental setup}
\label{tab:table_iii_experimental_setup}
\centering
\footnotesize
\renewcommand{\arraystretch}{1.40}
\begin{tabularx}{\columnwidth}{|l|X|}
\hline
\rowcolor{tableheader}
\multicolumn{1}{|c|}{\textcolor{white}{\textbf{Parameter}}} & \multicolumn{1}{c|}{\textcolor{white}{\textbf{Setting}}} \\ \hline

Problem sizes $h\times m$ & $8\times4$, $16\times8$, $32\times16$, $64\times32$ \\ \hline

Target load $\lambda_{\mathrm{sto}}=\lambda_{\mathrm{com}}$ & 0.75 \\ \hline

Training budget $E_{\max}$ / minimum & 3000 / 750 episodes \\ \hline

Greedy evaluation interval & 50 episodes \\ \hline

Stabilization window & 5 checkpoints with $\rho\ge0.999$ and range of $E\le0.01$ \\ \hline

Tuning & Seeds 100--104 ($16\times8$), 8 candidates per agent \\ \hline

Main evaluation seeds & 300--329 (30 instances per size) \\ \hline

Failure and shift seeds & 500--509 (10 instances) \\ \hline

Reward-ablation seeds & 700--709 (10 instances) \\ \hline

\ac{MILP} time limit / gap & 30 s / $10^{-6}$ \\ \hline

Network / parameters & 15--128--128--1 / 18,689 \\ \hline

Formulation constants & $\beta=0.5$, $\eta=0.25$, $\kappa=2$ \\ \hline
\end{tabularx}
\end{table}

We compare the agents with five heuristics and the \ac{MILP} in \eqref{eq:milp-objective}, using the definitions of~\cite{Asif008Paper}. Random feasible placement selects uniformly among the feasible \acp{VM}, and first-fit selects the feasible \ac{VM} with the lowest index. First-fit decreasing sorts the requests in decreasing order of their maximum normalized storage and compute demand and then applies first-fit \cite{BinPackingRef}. Best-fit selects the feasible \ac{VM} that minimizes the mean of its normalized residual storage and compute fractions after placement. The resource-wastage-aware heuristic, introduced in~\cite{Asif008Paper}, minimizes a score that combines this mean residual fraction, the absolute imbalance between the two fractions, and a penalty of $0.05$ for activating an inactive \ac{VM}. In contrast to these sequential methods, the \ac{MILP} optimizes all requests jointly, using the solver settings given in Section~\ref{Sec:ProblemFormulation}. The performance metrics follow~\cite{Asif008Paper}. In figures, for the sake of simplicity, Double \ac{DQN} is abbreviated as DDQN.

\begin{table}[ht!]
\caption{Selected hyperparameters of the \ac{DQN} and Double \ac{DQN} agents}
\label{tab:table_iv_selected_hyperparameters}
\centering
\footnotesize
\renewcommand{\arraystretch}{1.40}
\setlength{\tabcolsep}{4pt}
\begin{tabular}{|l|c|c|}
\hline
\rowcolor{tableheader}
\multicolumn{1}{|c|}{\textcolor{white}{\textbf{Parameter}}} & \textcolor{white}{\textbf{DQN}} & \textcolor{white}{\textbf{Double DQN}} \\ \hline

Learning rate $\alpha$ & $5\times10^{-4}$ & $2\times10^{-3}$ \\ \hline

Discount factor $\gamma$ & 0.99 & 0.95 \\ \hline

Initial / final $\varepsilon$ & 1.00 / 0.05 & 0.70 / 0.05 \\ \hline

Target-update interval $K$ (gradient steps) & 500 & 250 \\ \hline

Weight decay & $10^{-5}$ & $10^{-4}$ \\ \hline

Exploration-decay fraction$^{\dagger}$ & \multicolumn{2}{c|}{0.80} \\ \hline

Minibatch size $B$ / replay capacity $N$ & \multicolumn{2}{c|}{128 / 50,000} \\ \hline

Warm-up transitions $N_w$ & \multicolumn{2}{c|}{500} \\ \hline

Training interval $F$ (transitions) & \multicolumn{2}{c|}{4} \\ \hline

Optimizer / gradient-clip norm & \multicolumn{2}{c|}{AdamW / 10} \\ \hline

Huber threshold $\delta$ & \multicolumn{2}{c|}{1} \\ \hline

\multicolumn{3}{|l|}{\scriptsize $^{\dagger}$Fraction of training episodes over which $\varepsilon$ decays linearly.} \\ \hline

\end{tabular}
\end{table}

The experimental setup is summarized in Table~\ref{tab:table_iii_experimental_setup}. Hyperparameters are selected on five $16\times8$ instances generated from seeds 100--104 (Section~\ref{Subsec:Hyperparameters}). For the main benchmark, each of the four sizes is evaluated on $30$ independent instances generated from seeds 300--329. As discussed in Section~\ref{Subsec:WhyRL}, a separate policy is trained on each instance and evaluated greedily on that same instance; transfer to unseen instances is examined in Section~\ref{Subsec:Failure}. Seeds 500--509 and 700--709 are used for the failure and distribution-shift experiments and the reward ablation, respectively, and none for tuning.

\begin{table}[ht!]
\caption{Empirical stabilization of the agents at $16\times8$}
\label{tab:table_v_empirical_stabilization}
\centering
\footnotesize
\renewcommand{\arraystretch}{1.40}
\setlength{\tabcolsep}{4pt}
\begin{tabularx}{\columnwidth}{|X|c|c|c|}
\hline
\rowcolor{tableheader}
\multicolumn{1}{|c|}{\textcolor{white}{\textbf{Agent}}} &
\textcolor{white}{\textbf{Stable runs}} &
\textcolor{white}{\shortstack{\textbf{Window}\\\textbf{start}}} &
\textcolor{white}{\shortstack{\textbf{Detection}\\\textbf{episode}}} \\ \hline

DQN & 28/30 & 800 & 1000 \\ \hline

Double DQN & 28/30 & 650 & 850 \\ \hline

\multicolumn{4}{|p{\dimexpr\columnwidth-2\tabcolsep\relax}|}{\scriptsize Medians over stable runs (episodes). Detection occurs four 50-episode intervals after the window start.} \\ \hline

\end{tabularx}
\end{table}

Placement quality is measured by the placed-\ac{VNFC} fraction $\rho$ in \eqref{eq:placed_fraction010}. Resource efficiency is assessed by the efficiency cost $J_\beta$ in \eqref{eq:efficiency-cost}, the number of active \acp{VM} $U$, and the active-\ac{VM} residual wastage $W_{\mathrm{act}}$ in \eqref{eq:W2}. For the agents, we additionally report the greedy evaluation score $E$, the number of training episodes, stabilization, training time, and memory allocation. Means are reported with normal-approximation $95\%$ confidence intervals, $\bar{x}\pm1.96\,s/\sqrt{n}$, where $s$ is the sample standard deviation over $n$ independent instances; since $\rho$ is bounded by one, these intervals are approximate near the boundary. Differences between methods are evaluated on paired instances, and training time and peak allocation are reported as medians with interquartile ranges.

\begin{figure*}
\centering
\begin{tcolorbox}
\includegraphics[width=0.98\textwidth]{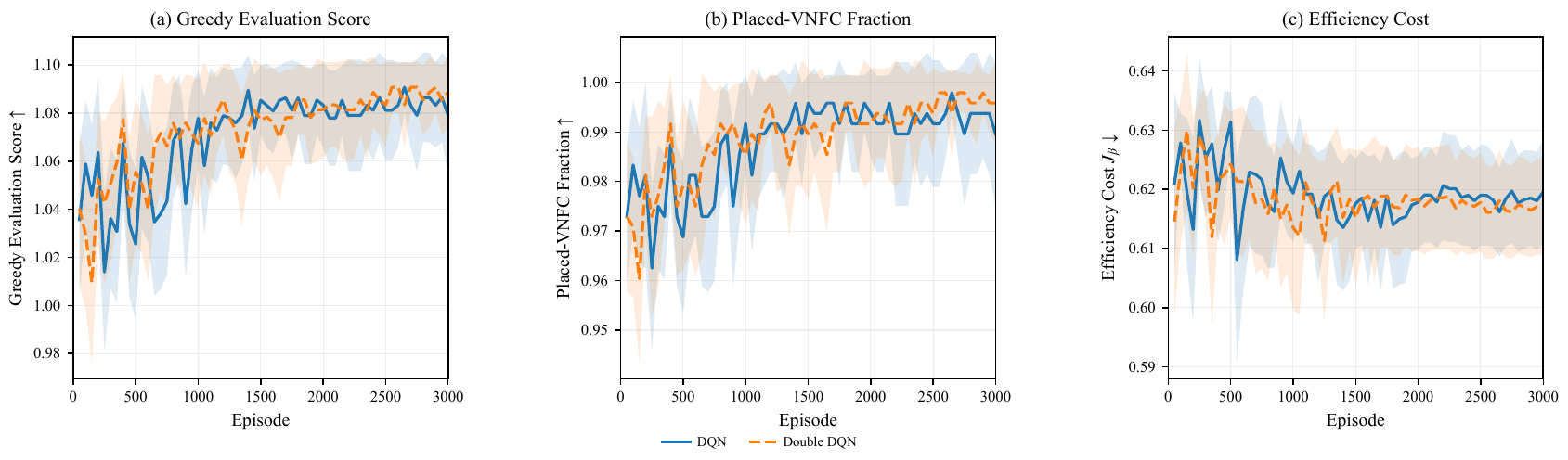}
\caption{Learning behavior of the two agents at $16\times8$: (a) greedy evaluation score $E$, which exceeds one only for complete mappings earning the efficiency bonus, (b) placed-\ac{VNFC} fraction $\rho$, and (c) efficiency cost $J_\beta$ of the greedy policy. Curves are means over the $30$ reference instances, and shaded bands are normal-approximation $95\%$ confidence intervals. For runs that stop early, the last evaluated value is carried forward.}
\label{fig:learning_behavior010}
\end{tcolorbox}
\vspace{-4.8mm}
\end{figure*}

For the agents, decision time includes action-feature construction and greedy action selection; for the heuristics, it covers the \ac{VM} selection rule. Both exclude environment updates. For the \ac{MILP}, which solves the batch jointly, the reported per-\ac{VNFC} time is the solver time divided by $h$. Peak allocation is measured with Python's \texttt{tracemalloc}, which traces Python-level and NumPy allocations, such as the replay buffer, but not memory allocated internally by PyTorch; it therefore does not represent the total process-resident memory.

\subsection{Hyperparameter Selection}\label{Subsec:Hyperparameters}
Each agent is tuned over eight candidate configurations, using the same five $16\times8$ instances. The candidates are ranked lexicographically by the divergence rate, i.e., the fraction of runs with numerically unstable training (lower is better), followed by the mean placed-\ac{VNFC} fraction (higher), the mean greedy evaluation score (higher), the mean efficiency cost (lower), and the mean training time (lower). The selected hyperparameters, listed in Table~\ref{tab:table_iv_selected_hyperparameters}, are fixed before the main benchmark, the failure analysis, and the reward ablation, and are used unchanged for all problem sizes.

Both agents use the architecture described in Section~\ref{Subsec:RLSolution}, with $18{,}689$ trainable parameters per online network, independent of the number of \acp{VM}. Their tuned hyperparameters, however, differ: \ac{DQN} uses $\gamma=0.99$, a learning rate of $5\times10^{-4}$, and a target-update interval of $500$ gradient steps, whereas Double \ac{DQN} uses $\gamma=0.95$, a learning rate of $2\times10^{-3}$, and an interval of $250$ gradient steps. The comparison evaluates the agents under independently tuned configurations.

\subsection{Learning Behavior and Empirical Stabilization}\label{Subsec:LearningBehavior010}
Figure~\ref{fig:learning_behavior010} shows the greedy evaluation score, the placed-\ac{VNFC} fraction, and the efficiency cost during training, averaged over the $30$ reference $16\times8$ instances. Both agents fluctuate strongly at early checkpoints, after which their mean evaluation scores and placed fractions improve and vary less. Scores above one correspond to complete mappings that earn the efficiency bonus in \eqref{eq:evaluation_score010}. The efficiency cost $J_\beta$ remains within $[0.605, 0.635]$ throughout, indicating that training mainly improves feasibility rather than efficiency. If a run terminates early, its last evaluated value is carried forward in the plot only, without implying further training.

To distinguish observed stabilization from theoretical convergence, we apply the operational stopping criterion of~\cite{Asif008Paper}. Greedy evaluations are performed every $50$ episodes, and the stabilization test is enabled after at least $750$ episodes. A run is considered stabilized when its five latest checkpoints all satisfy $\rho\ge0.999$, i.e., yield complete mappings, and the range of their evaluation scores $E$ is at most $0.01$. Runs that do not meet this rule continue up to $E_{\max}=3000$ episodes. As summarized in Table~\ref{tab:table_v_empirical_stabilization}, both agents stabilize in $28$ of the $30$ runs, and Double \ac{DQN} does so about $150$ episodes earlier in median. Since the two agents use different hyperparameters, this difference cannot be attributed to the target construction alone, and neither the stopping rule nor the learning curves certify convergence to the optimal action-value function.

\subsection{Main Benchmark at \texorpdfstring{$16\times8$}{16x8}}\label{Subsec:PlacementQuality010}
\begin{figure*}[t]
\centering
\begin{tcolorbox}
\includegraphics[width=0.98\textwidth]{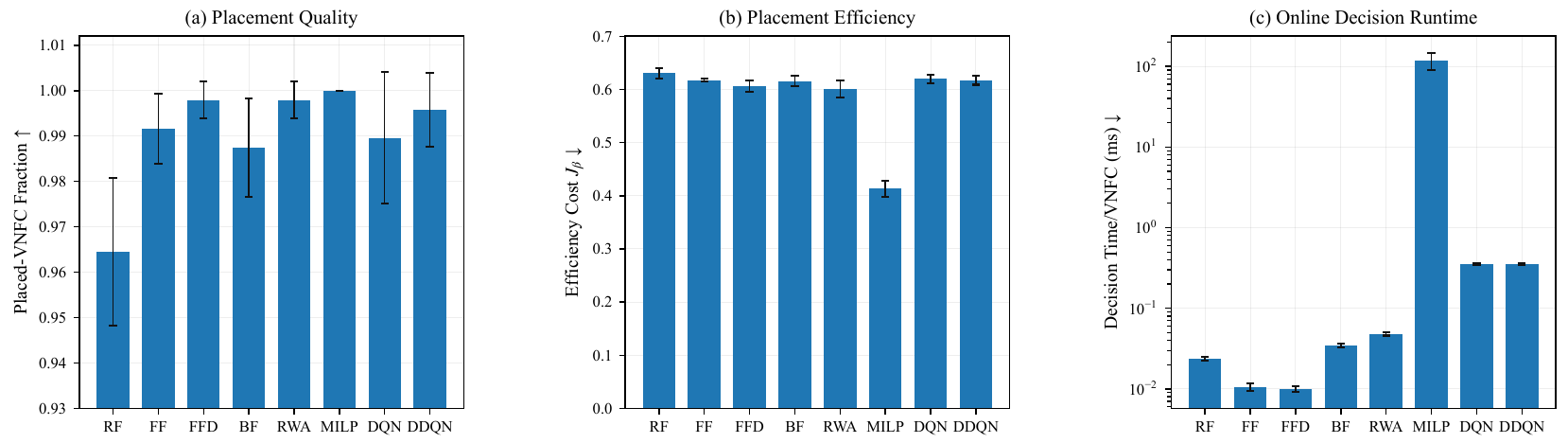}
\caption{Main comparison at $16\times8$: (a) placed-\ac{VNFC} fraction $\rho$, (b) efficiency cost $J_\beta$, and (c) decision time per \ac{VNFC} on a logarithmic scale, where the \ac{MILP} time is the batch solve time divided by $h$. RF: random feasible placement; FF: first-fit; FFD: first-fit decreasing; BF: best-fit; RWA: resource-wastage-aware heuristic; DDQN: Double \ac{DQN}. Error bars show normal-approximation $95\%$ confidence intervals over $30$ instances.}
\label{fig:main_comparison010}
\end{tcolorbox}
\vspace{-4.8mm}
\end{figure*}

The main benchmark compares the two agents and the six baselines on the reference $16\times8$ setup. Figure~\ref{fig:main_comparison010} summarizes placement quality, efficiency, and online decision time, and Table~\ref{tab:table_vi_main_16x8} lists the corresponding values with $95\%$ confidence intervals. Both agents achieve high mean placed fractions, $0.990\pm0.014$ for \ac{DQN} and $0.996\pm0.008$ for Double \ac{DQN}. However, first-fit decreasing and the resource-wastage-aware heuristic match or slightly exceed them, both reaching $0.998\pm0.004$, with overlapping confidence intervals, and the \ac{MILP} places every \ac{VNFC} in all $30$ instances.

The differences become clearer in resource efficiency. \ac{DQN} and Double \ac{DQN} obtain mean costs of $J_\beta=0.620$ and $0.617$, respectively, compared with $0.606$ for first-fit decreasing and $0.601$ for the resource-wastage-aware heuristic, although the confidence intervals overlap. Their active-\ac{VM} wastage of $0.244$ and $0.239$ is likewise slightly higher than the $0.221$ of both heuristics. The \ac{MILP} attains a substantially lower cost, $J_\beta=0.414\pm0.015$, and a wastage of only $0.048\pm0.011$, because it optimizes the batch jointly and consolidates the \acp{VNFC} onto fewer, more fully utilized \acp{VM}. Hence, complete placement does not imply efficient resource utilization: the agents achieve high completion but neither match the \ac{MILP} in efficiency nor consistently improve on the strongest heuristics.

\begin{table*}[t]
\caption{Results at $16\times8$: means $\pm$ normal-approximation $95\%$ confidence half-widths over $30$ instances}
\label{tab:table_vi_main_16x8}
\centering
\footnotesize
\renewcommand{\arraystretch}{1.40}
\setlength{\tabcolsep}{5pt}
\begin{tabular}{|l|c|c|c|c|}
\hline
\rowcolor{tableheader}
\multicolumn{1}{|c|}{\textcolor{white}{\textbf{Method}}} &
\textcolor{white}{\textbf{Placed-VNFC fraction $\boldsymbol{\rho}$}} &
\textcolor{white}{\textbf{Efficiency cost $\boldsymbol{J_\beta}$}} &
\textcolor{white}{\textbf{Active-VM wastage $\boldsymbol{W_{\mathrm{act}}}$}} &
\textcolor{white}{\textbf{Decision time (ms/VNFC)}} \\ \hline

Random feasible placement & $0.965 \pm 0.016$ & $0.630 \pm 0.010$ & $0.265 \pm 0.014$ & $0.023 \pm 0.001$ \\ \hline

First-fit & $0.992 \pm 0.008$ & $0.617 \pm 0.003$ & $0.235 \pm 0.006$ & $0.011 \pm 0.001$ \\ \hline

First-fit decreasing & $0.998 \pm 0.004$ & $0.606 \pm 0.011$ & $0.221 \pm 0.011$ & $0.010 \pm 0.001$ \\ \hline

Best-fit & $0.988 \pm 0.011$ & $0.615 \pm 0.010$ & $0.235 \pm 0.012$ & $0.034 \pm 0.002$ \\ \hline

Resource-wastage-aware heuristic & $0.998 \pm 0.004$ & $0.601 \pm 0.016$ & $0.221 \pm 0.015$ & $0.048 \pm 0.002$ \\ \hline

\ac{MILP}$^{\ast}$ & $1.000 \pm 0.000$ & $0.414 \pm 0.015$ & $0.048 \pm 0.011$ & $118.29 \pm 28.04^{\dagger}$ \\ \hline\hline

\ac{DQN} & $0.990 \pm 0.014$ & $0.620 \pm 0.009$ & $0.244 \pm 0.010$ & $0.354 \pm 0.007$ \\ \hline

Double \ac{DQN} & $0.996 \pm 0.008$ & $0.617 \pm 0.008$ & $0.239 \pm 0.009$ & $0.352 \pm 0.007$ \\ \hline

\multicolumn{5}{|l|}{\scriptsize $^{\ast}$All $30$ solutions certified optimal at this size. $^{\dagger}$Batch solve time divided by $h$. Intervals are approximate near the bound $\rho=1$.} \\ \hline

\end{tabular}
\vspace{-4.8mm}
\end{table*}

In terms of decision time, \ac{DQN} and Double \ac{DQN} require $0.354$ and $0.352$~ms per \ac{VNFC}, about $35$ times longer than first-fit decreasing ($0.010$~ms) and $7$ times longer than the resource-wastage-aware heuristic ($0.048$~ms). The \ac{MILP} requires $118.29$~ms per \ac{VNFC}, i.e., about $1.9$~s per $16\times8$ batch, and all $30$ solutions at this size are certified optimal. Once trained, the agents decide more than two orders of magnitude faster than the \ac{MILP} but remain slower than the heuristics.

\subsection{Placement Quality and Computational Scaling}\label{Subsec:Scalability010}
\begin{figure*}[t]
\centering
\begin{tcolorbox}
\includegraphics[width=0.98\textwidth]{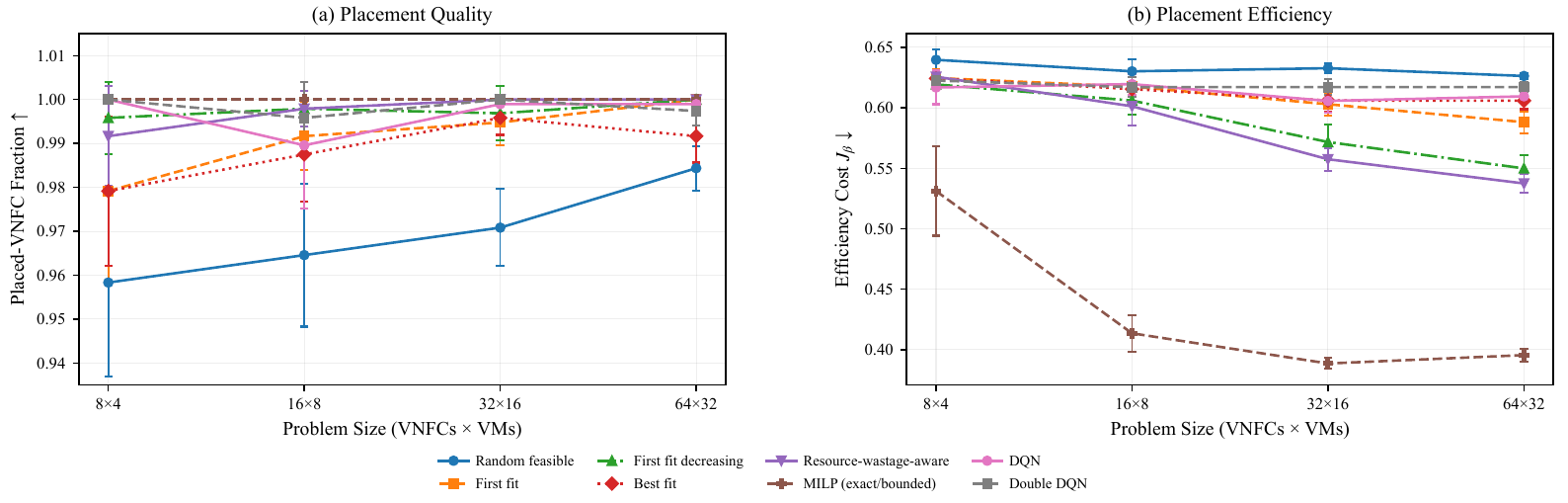}
\caption{Scalability across the four problem sizes: (a) mean placed-\ac{VNFC} fraction $\rho$ and (b) mean efficiency cost $J_\beta$. Error bars show normal-approximation $95\%$ confidence intervals over $30$ instances per size. The agents are trained separately on each instance. At $32\times16$ and $64\times32$, the \ac{MILP} values are mostly time-bounded incumbents rather than certified optima.}
\label{fig:controlled_scalability010}
\end{tcolorbox}
\vspace{-4.8mm}
\end{figure*}

We next examine how placement quality and computational cost change across the four problem sizes. Figure~\ref{fig:controlled_scalability010} reports the mean placed-\ac{VNFC} fraction and efficiency cost of all eight approaches. Both agents maintain high completion across all sizes, with mean placed fractions from $0.990$ to $1.000$ for \ac{DQN} and from $0.996$ to $1.000$ for Double \ac{DQN}. At $8\times4$, both place all requests in all $30$ instances, and at $64\times32$ they reach $0.999\pm0.001$ and $0.997\pm0.003$, respectively. This suggests that the action-conditioned representation remains effective within the studied sizes. Since a new model is trained for each instance, however, this experiment does not establish transfer to unseen problem sizes, which is examined in Section~\ref{Subsec:Failure}.

\begin{figure*}[t]
\centering
\begin{tcolorbox}
\includegraphics[width=0.98\textwidth]{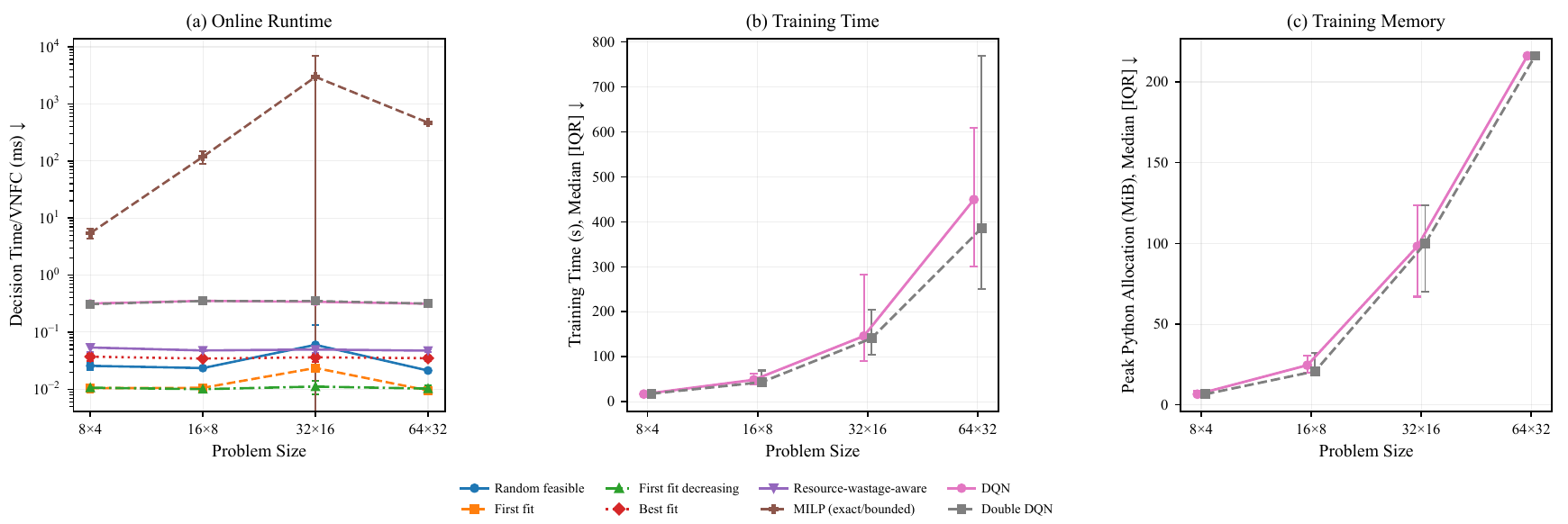}
\caption{Computational cost across problem sizes: (a) mean decision time per \ac{VNFC} on a logarithmic scale, with the \ac{MILP} time given as batch solve time divided by $h$, (b) median training time of the agents, and (c) median peak allocation traced by \texttt{tracemalloc} during training, which excludes memory allocated internally by PyTorch. Error bars show $95\%$ confidence intervals for decision times and interquartile ranges for training time and allocation.}
\label{fig:computational_scaling010}
\end{tcolorbox}
\vspace{-4.8mm}
\end{figure*}

Near-complete placement does not eliminate differences in resource utilization. At $64\times32$, \ac{DQN} achieves $J_\beta=0.609\pm0.006$ and Double \ac{DQN} $0.617\pm0.004$, while first-fit decreasing and the resource-wastage-aware heuristic still attain lower mean costs. The mean paired difference in $J_\beta$ (Double \ac{DQN} minus \ac{DQN}) favors \ac{DQN} at $32\times16$ and $64\times32$, at $0.011$ $[0.001, 0.022]$ and $0.008$ $[0.001, 0.014]$, respectively. Since these unadjusted intervals stem from exploratory comparisons across several configurations and independently tuned agents, they do not show that \ac{DQN} is generally more efficient. The \ac{MILP} attains the lowest mean cost at every size, but its certification status changes markedly: all $30$ solutions are certified optimal at $8\times4$ and $16\times8$, compared with $1$ at $32\times16$ and none at $64\times32$. Its results at the two larger sizes are time-bounded incumbents rather than certified optima.

\begin{table*}[ht!]
\caption{Computational cost of the agents across problem sizes: decision times are means $\pm$ $95\%$ confidence half-widths; training times and peak allocations are medians [Q1, Q3]}
\label{tab:table_vii_neural_scalability_computational_cost}
\centering
\footnotesize
\renewcommand{\arraystretch}{1.40}
\setlength{\tabcolsep}{5pt}
\begin{tabular}{|c|l|c|c|c|}
\hline
\rowcolor{tableheader}
\textcolor{white}{\textbf{Size}} &
\multicolumn{1}{c|}{\textcolor{white}{\textbf{Agent}}} &
\textcolor{white}{\textbf{Decision time (ms/VNFC)}} &
\textcolor{white}{\textbf{Training time (s)}} &
\textcolor{white}{\textbf{Peak allocation (MiB)}} \\ \hline
\multirow{2}{*}{$8\times4$}   & \ac{DQN}        & $0.318 \pm 0.012$ & $16.9$ [$16.1$, $20.8$]    & $6.6$ [$6.5$, $8.5$] \\ \cline{2-5}
                              & Double \ac{DQN} & $0.311 \pm 0.012$ & $17.8$ [$17.1$, $20.1$]    & $6.6$ [$6.5$, $7.6$] \\ \hline
                              
\multirow{2}{*}{$16\times8$}  & \ac{DQN}        & $0.354 \pm 0.007$ & $48.6$ [$39.0$, $61.6$]    & $24.6$ [$19.8$, $30.3$] \\ \cline{2-5}
                              & Double \ac{DQN} & $0.352 \pm 0.007$ & $43.3$ [$40.0$, $69.3$]    & $20.8$ [$18.5$, $32.3$] \\ \hline
                              
\multirow{2}{*}{$32\times16$} & \ac{DQN}        & $0.340 \pm 0.007$ & $145.8$ [$90.2$, $282.0$]  & $98.2$ [$67.0$, $123.6$] \\ \cline{2-5}
                              & Double \ac{DQN} & $0.351 \pm 0.005$ & $140.8$ [$104.2$, $203.5$] & $99.9$ [$70.1$, $123.6$] \\ \hline
                              
\multirow{2}{*}{$64\times32$} & \ac{DQN}        & $0.315 \pm 0.013$ & $449.2$ [$300.5$, $609.1$] & $216.1$ [$216.1$, $216.2$] \\ \cline{2-5}
                              & Double \ac{DQN} & $0.317 \pm 0.010$ & $385.5$ [$249.9$, $769.0$] & $216.1$ [$216.1$, $216.2$] \\ \hline
\end{tabular}
\vspace{-4.8mm}
\end{table*}

Figure~\ref{fig:computational_scaling010} shows the computational cost across sizes. The agents' decision times remain between $0.31$ and $0.35$~ms per \ac{VNFC} over the evaluated range, as the candidate \acp{VM} are scored in a single batched forward pass whose fixed overhead dominates at these sizes. This does not imply size-independent complexity, since every candidate \ac{VM} still requires a value evaluation. The heuristics remain faster for online decisions, whereas the \ac{MILP} is orders of magnitude more expensive. Its per-\ac{VNFC} time is non-monotonic because, once the solver reaches the $30$~s limit at the larger sizes, the capped batch time is divided by a growing number of requests $h$.

The agents' main cost lies in training. In Table~\ref{tab:table_vii_neural_scalability_computational_cost}, median training time grows from $16.9$~s at $8\times4$ to $449.2$~s at $64\times32$ for \ac{DQN} and from $17.8$~s to $385.5$~s for Double \ac{DQN}, with broad interquartile ranges at the larger sizes. Since a policy is trained per instance, this cost exceeds the $30$~s \ac{MILP} limit by more than an order of magnitude at $64\times32$. The median peak traced allocation rises from $6.6$~MiB to $216.1$~MiB for both agents, although each online network retains the same $18{,}689$ parameters. This growth stems from the replay buffer: each transition stores the successor features $\Phi'$ of all $m$ \acp{VM}, and at $64\times32$ the buffer reaches its capacity of $50{,}000$ transitions, which explains the nearly constant allocation across runs.

These measurements distinguish representation size from computational scalability: a fixed number of network parameters does not imply constant training or memory cost, and therefore does not by itself establish superior deployment efficiency. Overall, neural inference avoids repeated \ac{MILP} optimization, but at a substantial offline training cost and without an online-latency advantage over strong heuristics.

\subsection{Failure Conditions and Distribution Shifts}\label{Subsec:Failure}
\begin{figure*}[t]
\centering
\begin{tcolorbox}
\includegraphics[width=0.98\textwidth] {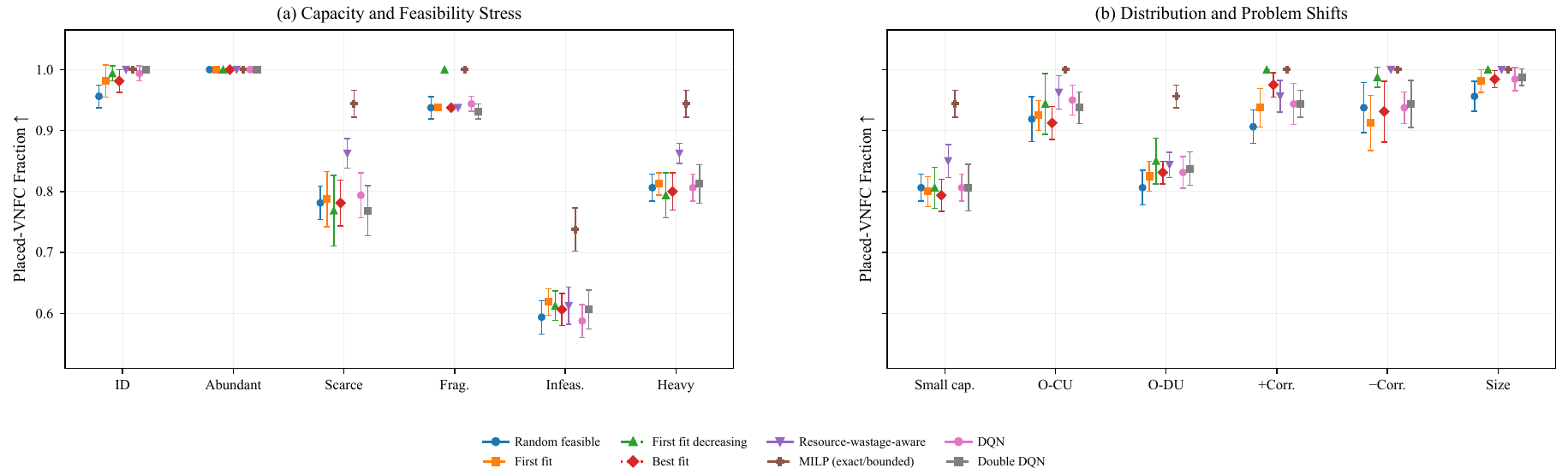}
\caption{Placed-\ac{VNFC} fraction under (a) capacity and feasibility stress and (b) changed demand profiles, correlations, and problem size, averaged over ten instances (seeds 500--509). ID: in-distribution reference; Frag.: fragmentation instance; Infeas.: globally infeasible load; Small cap.: reduced \ac{VM} capacities; O-CU/O-DU: \ac{O-CU}-/\ac{O-DU}-heavy profiles; $\pm$Corr.: positively/negatively correlated demands; Size: unseen problem size. The agents are trained on the ID instance of each seed and evaluated without retraining. Error bars show normal-approximation $95\%$ confidence intervals.}
\label{fig:failure_shift010}
\end{tcolorbox}
\vspace{-4.8mm}
\end{figure*}

To identify conditions under which the methods lose completion performance, we evaluate capacity stress and controlled distribution shifts on ten $16\times8$ instances generated from seeds 500--509. For each seed, both agents are trained on the corresponding instance, and the trained policies are then evaluated without retraining in all scenarios; the in-distribution reference (ID) is the training instance itself. Figure~\ref{fig:failure_shift010} and Table~\ref{tab:table_viii_failure_shift010} cover the ID reference, abundant and scarce resources, a fragmentation instance (Frag.), global infeasibility (Infeas.), heavier demands, smaller \ac{VM} capacities (Small cap.), \ac{O-CU}- and \ac{O-DU}-heavy request profiles, positively and negatively correlated demands ($+$Corr., $-$Corr.), and an unseen problem size of $32\times16$. The reference, abundant, scarce, and infeasible cases use target loads $\lambda_{\mathrm{sto}}=\lambda_{\mathrm{com}}$ of $0.75$, $0.50$, $0.95$, and $1.20$, respectively. Heavier demands scale all demands by $1.25$, and small capacities scale all \ac{VM} capacities by $0.80$. The \ac{O-CU}-heavy case multiplies the compute demands of the \ac{O-CU} components by $1.50$, and the \ac{O-DU}-heavy case the storage demands of the \ac{O-DU} components, leaving all other demands unchanged. For $+$Corr., the storage and compute demands are each sorted in ascending order and then paired; for $-$Corr., the storage demands are sorted in ascending and the compute demands in descending order. The fragmentation case is a globally feasible but order-sensitive fixed instance.

Under abundant resources, all approaches place every request. In contrast, scarce resources reduce the mean placed fraction to $0.794$ for \ac{DQN} and $0.769$ for Double \ac{DQN}, heavier unseen demands to $0.806$ and $0.812$, and smaller \ac{VM} capacities to $0.806$ for both agents. In these three cases, the resource-wastage-aware heuristic retains higher fractions ($0.850$--$0.863$), and the \ac{MILP} reaches $0.944$, mainly because it optimizes the whole batch jointly. The globally infeasible case serves as a consistency check, since complete placement is impossible when capacity is insufficient: the agents' mean fractions fall to $0.588$ and $0.606$, compared with $0.738$ for the \ac{MILP}. This reduction reflects physical infeasibility, method-specific allocation behavior, and, for the agents, the transfer of policies trained under the reference conditions, and should not be interpreted as a learning failure alone.

The fragmentation instance highlights the importance of request order and resource fit: first-fit decreasing and the \ac{MILP} place all requests, whereas \ac{DQN} and Double \ac{DQN} reach $0.944$ and $0.931$, on par with the remaining heuristics ($0.938$). The \ac{O-CU}-heavy and correlation scenarios cause moderate changes, whereas the \ac{O-DU}-heavy scenario, which scales storage demands, reduces all non-\ac{MILP} methods to $0.81$--$0.85$. Hence, performance under a distribution shift depends on the remaining feasibility margins rather than on the existence of a shift. In the unseen-size test, the trained networks are applied to $32\times16$ and reach placed fractions of $0.984$ for \ac{DQN} and $0.988$ for Double \ac{DQN}, close to their in-distribution level but below first-fit decreasing, the resource-wastage-aware heuristic, and the \ac{MILP}, which place all requests. Unlike the scalability experiment, no new network is trained here. These results characterize behavior under selected changes but do not establish broad out-of-distribution robustness.

\begin{table*}[t]
\caption{Mean placed-\ac{VNFC} fraction under failure conditions and distribution shifts over ten instances (seeds 500--509); abbreviations and confidence intervals as in Figure~\ref{fig:failure_shift010}}
\label{tab:table_viii_failure_shift010}
\centering
\footnotesize
\renewcommand{\arraystretch}{1.40}
\setlength{\tabcolsep}{5pt}
\begin{tabular}{|l|c|c|c|c|c|c|}
\hline
\rowcolor{tableheader}
\textcolor{white}{\textbf{Method}} & \textcolor{white}{\textbf{ID}} & \textcolor{white}{\textbf{Abundant}} & \textcolor{white}{\textbf{Scarce}} & \textcolor{white}{\textbf{Frag.}} & \textcolor{white}{\textbf{Infeas.}} & \textcolor{white}{\textbf{Heavy}} \\ \hline
Random feasible placement        & 0.956 & 1.000 & 0.781 & 0.938 & 0.594 & 0.806 \\ \hline
First-fit                        & 0.981 & 1.000 & 0.787 & 0.938 & 0.619 & 0.812 \\ \hline
First-fit decreasing             & 0.994 & 1.000 & 0.769 & 1.000 & 0.613 & 0.794 \\ \hline
Best-fit                         & 0.981 & 1.000 & 0.781 & 0.938 & 0.606 & 0.800 \\ \hline
Resource-wastage-aware heuristic & 1.000 & 1.000 & 0.863 & 0.938 & 0.613 & 0.863 \\ \hline
\ac{MILP}$^{\ast}$               & 1.000 & 1.000 & 0.944 & 1.000 & 0.738 & 0.944 \\ \hline\hline
\ac{DQN}                         & 0.994 & 1.000 & 0.794 & 0.944 & 0.588 & 0.806 \\ \hline
Double \ac{DQN}                  & 1.000 & 1.000 & 0.769 & 0.931 & 0.606 & 0.812 \\ \hline
\rowcolor{tableheader}
\textcolor{white}{\textbf{Method}} & \textcolor{white}{\textbf{Small cap.}} & \textcolor{white}{\textbf{O-CU}} & \textcolor{white}{\textbf{O-DU}} & \textcolor{white}{\textbf{$+$Corr.}} & \textcolor{white}{\textbf{$-$Corr.}} & \textcolor{white}{\textbf{Size}} \\ \hline
Random feasible placement        & 0.806 & 0.919 & 0.806 & 0.906 & 0.938 & 0.956 \\ \hline
First-fit                        & 0.800 & 0.925 & 0.825 & 0.938 & 0.912 & 0.981 \\ \hline
First-fit decreasing             & 0.806 & 0.944 & 0.850 & 1.000 & 0.988 & 1.000 \\ \hline
Best-fit                         & 0.794 & 0.912 & 0.831 & 0.975 & 0.931 & 0.984 \\ \hline
Resource-wastage-aware heuristic & 0.850 & 0.963 & 0.844 & 0.956 & 1.000 & 1.000 \\ \hline
\ac{MILP}$^{\ast}$               & 0.944 & 1.000 & 0.956 & 1.000 & 1.000 & 1.000 \\ \hline\hline
\ac{DQN}                         & 0.806 & 0.950 & 0.831 & 0.944 & 0.938 & 0.984 \\ \hline
Double \ac{DQN}                  & 0.806 & 0.938 & 0.838 & 0.944 & 0.944 & 0.988 \\ \hline
\multicolumn{7}{|l|}{\scriptsize $^{\ast}$Solved with a $30$~s time limit; solutions are not necessarily certified optimal.} \\ \hline
\end{tabular}
\vspace{-4.8mm}
\end{table*}

\subsection{Reward Ablation}\label{Subsec:RewardAblation010}
\begin{figure*}[t]
\begin{tcolorbox}
\centering
\includegraphics[width=0.98\textwidth]{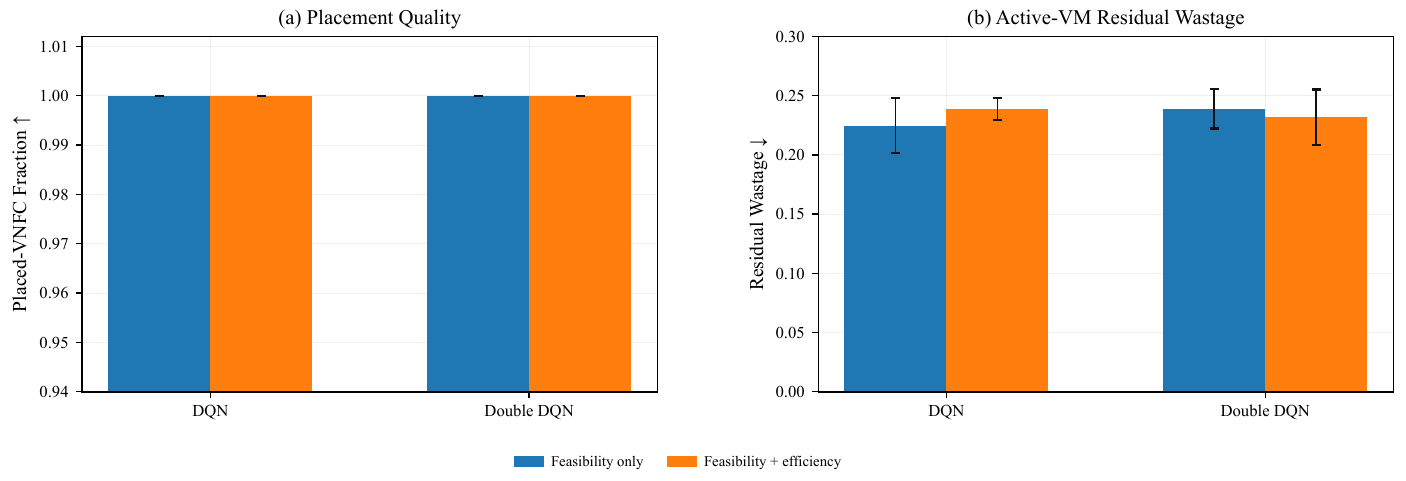}
\caption{Reward ablation at $16\times8$ with feasibility-only and feasibility-plus-efficiency rewards: (a) placed-\ac{VNFC} fraction, which equals one for all configurations, and (b) active-\ac{VM} residual wastage $W_{\mathrm{act}}$. Error bars show normal-approximation $95\%$ confidence intervals over ten instances (seeds 700--709).}
\label{fig:reward_ablation010}
\end{tcolorbox}
\vspace{-4.8mm}
\end{figure*}

The reward in \eqref{eq:revised-reward} combines a basic placement reward with an efficiency bonus that is awarded only when all requests are placed. To examine whether this bonus improves the learned resource allocation, we compare the full reward with a feasibility-only variant without the efficiency bonus on ten $16\times8$ instances generated from seeds 700--709. The architecture and hyperparameters are unchanged, so that only the reward differs. Figure~\ref{fig:reward_ablation010} and Table~\ref{tab:table_ix_reward_ablation010} summarize the placed fractions, efficiency costs, and active-\ac{VM} wastage.

\begin{table*}[t]
\caption{Reward ablation at $16\times8$: means $\pm$ normal-approximation $95\%$ confidence half-widths over ten instances (seeds 700--709)}
\label{tab:table_ix_reward_ablation010}
\centering
\footnotesize
\renewcommand{\arraystretch}{1.15}
\setlength{\tabcolsep}{5pt}
\begin{tabular}{|l|l|c|c|c|}
\hline
\rowcolor{tableheader}
\multicolumn{1}{|c|}{\textcolor{white}{\textbf{Agent}}} &
\multicolumn{1}{c|}{\textcolor{white}{\textbf{Reward}}} &
\textcolor{white}{\textbf{Placed-VNFC fraction $\boldsymbol{\rho}$}} &
\textcolor{white}{\textbf{Efficiency cost $\boldsymbol{J_\beta}$}} &
\textcolor{white}{\textbf{Active-VM wastage $\boldsymbol{W_{\mathrm{act}}}$}} \\ \hline
\multirow{2}{*}{\ac{DQN}}        & Feasibility only$^{\ast}$ & $1.000 \pm 0.000$ & $0.598 \pm 0.030$ & $0.225 \pm 0.023$ \\ \cline{2-5}
                                 & Feasibility + efficiency  & $1.000 \pm 0.000$ & $0.619 \pm 0.005$ & $0.239 \pm 0.009$ \\ \hline
\multirow{2}{*}{Double \ac{DQN}} & Feasibility only$^{\ast}$ & $1.000 \pm 0.000$ & $0.612 \pm 0.022$ & $0.239 \pm 0.017$ \\ \cline{2-5}
                                 & Feasibility + efficiency  & $1.000 \pm 0.000$ & $0.609 \pm 0.025$ & $0.232 \pm 0.024$ \\ \hline
\multicolumn{5}{|l|}{\scriptsize $^{\ast}$Reward \eqref{eq:revised-reward} with $\eta=0$.} \\ \hline
\end{tabular}
\vspace{-4.8mm}
\end{table*}

Both agents place all requests under both reward definitions, so the terminal bonus has no effect on completion in these instances. For \ac{DQN}, the bonus increases the mean efficiency cost from $0.598\pm0.030$ to $0.619\pm0.005$ and the active-\ac{VM} wastage from $0.225$ to $0.239$, although it markedly reduces the variability across instances. For Double \ac{DQN}, the effect is smaller and in the opposite direction, with the cost decreasing from $0.612\pm0.022$ to $0.609\pm0.025$ and the wastage from $0.239$ to $0.232$. For both agents, these differences lie within the confidence intervals. Hence, the terminal efficiency bonus with $\eta=0.25$ does not yield a consistent improvement in resource efficiency. We retain it for consistency with~\cite{Asif008Paper}; a likely reason for its limited effect is that a bonus granted only once per episode provides a sparse learning signal, which motivates denser efficiency rewards in future work.

\subsection{Comparison with Tabular and Linear Methods}\label{Subsec:ComparisonPrior}
We compare \ac{DQN} and Double \ac{DQN} with the tabular and linear variants of SARSA and $Q$-learning investigated in our previous work~\cite{Asif008Paper}. Both studies use the same placement objective, constraints, performance metrics, and evaluation instances (seeds 300--329); the results of the tabular and linear methods are taken from~\cite{Asif008Paper}, and Table~\ref{tab:comparison008010} reports them for the reference $16\times8$ setup.

\begin{table}[t]
\caption{Comparison with the tabular and linear methods of~\cite{Asif008Paper} (results taken from~\cite{Asif008Paper}, same instances) at $16\times8$: means $\pm$ normal-approximation $95\%$ confidence half-widths over $30$ instances}
\label{tab:comparison008010}
\centering
\footnotesize
\renewcommand{\arraystretch}{1.15}
\setlength{\tabcolsep}{4pt}
\begin{tabularx}{\columnwidth}{|X|c|c|c|}
\hline
\rowcolor{tableheader}
\multicolumn{1}{|c|}{\textcolor{white}{\textbf{Method}}} &
\textcolor{white}{\textbf{Parameters}} &
\textcolor{white}{$\boldsymbol{\rho}$} &
\textcolor{white}{$\boldsymbol{J_\beta}$} \\ \hline
Tabular SARSA          & size-dep. & $0.996 \pm 0.008$ & $0.621 \pm 0.004$ \\ \hline
Tabular $Q$-learning   & size-dep. & $0.992 \pm 0.013$ & $0.618 \pm 0.008$ \\ \hline
Linear SARSA           & $15$      & $0.992 \pm 0.010$ & $0.617 \pm 0.010$ \\ \hline
Linear $Q$-learning    & $15$      & $0.990 \pm 0.010$ & $0.618 \pm 0.010$ \\ \hline\hline
\ac{DQN}               & $18{,}689$ & $0.990 \pm 0.014$ & $0.620 \pm 0.009$ \\ \hline
Double \ac{DQN}        & $18{,}689$ & $0.996 \pm 0.008$ & $0.617 \pm 0.008$ \\ \hline
\end{tabularx}
\end{table}

All six methods achieve high placed fractions, from $0.990$ to $0.996$, and nearly identical efficiency costs, from $0.617$ to $0.621$. Double \ac{DQN} matches the placed fraction of tabular SARSA, and \ac{DQN} that of linear $Q$-learning. These results therefore show no consistent improvement in placement quality or efficiency attributable to neural function approximation.

The methods differ mainly in their learned representations: the tabular methods store size-dependent state-action values, whereas the linear and neural methods use $15$ and $18{,}689$ parameters, respectively, independent of the number of candidate \acp{VM} (Table~\ref{tab:comparison008010}). The neural representation thus adds training and inference cost without improving placement quality or efficiency on this static problem, and the comparison does not establish the superiority of either approach. Its potential advantages lie rather in transfer across instances and in dynamic settings, as indicated by the zero-shot results in Section~\ref{Subsec:Failure}.

\subsection{Limitations and Scope of the Evaluation}\label{Subsec:Limitations010}
The results should be interpreted within the scope of the simulated setup. First, the experiments consider static, ordered batches of \acp{VNFC}; they therefore do not establish performance in dynamic environments or in a deployed \ac{O-Cloud} controlled by an operational \ac{Non-RT RIC}. Second, the model captures storage and compute capacities only, without O-RAN-specific constraints such as fronthaul or F1 latency budgets and placement affinities, and the instances are synthetic. Third, the evaluated sizes extend only to $64\times32$ and do not establish behavior at larger scales. Fourth, the main benchmark trains and evaluates each policy on the same instance, so it measures instance-specific optimization rather than generalization; the zero-shot analysis covers only a selected set of stress and shift scenarios. Fifth, the hyperparameters of both agents were tuned independently and only at $16\times8$, which limits the attribution of their differences to the target construction. Finally, the decision times exclude offline training, and the traced allocations do not represent the full process memory.


\section{Concluding Remarks and Future Outlook}\label{Sec:Conclusion}
In this paper, we extended \texttt{SliceMapper}~\cite{Asif008Paper} by investigating \ac{DQN} and Double \ac{DQN} agents for the capacity-constrained mapping of \ac{O-CU} and \ac{O-DU} \acp{VNFC} onto \acp{VM} in \acp{O-Cloud}, and evaluated them against five heuristics and a time-bounded \ac{MILP}. At $16\times8$, \ac{DQN} and Double \ac{DQN} achieved mean placed-\ac{VNFC} fractions of $0.990\pm0.014$ and $0.996\pm0.008$, and both satisfied the empirical stabilization criterion in $28$ of $30$ runs. However, first-fit decreasing and the resource-wastage-aware heuristic matched or slightly exceeded their placement quality at lower efficiency cost and lower decision time, while the \ac{MILP} attained the lowest efficiency cost at a much higher runtime. Across the investigated sizes, the agents maintained high placed fractions, but their training time and memory grew substantially; with per-instance training, this cost exceeds a single \ac{MILP} solve at the larger sizes. Neither Double \ac{DQN} over \ac{DQN} nor the terminal efficiency bonus yielded a consistent improvement, and the agents achieved placement quality and efficiency comparable to the tabular and linear methods of~\cite{Asif008Paper}. The results highlight the trade-offs of neural placement and do not establish its superiority over conventional methods for the static problem considered.

Future work will address the settings in which learned policies may pay off. First, we will train policies across distributions of instances and evaluate them on unseen instances and sizes, so that the inference advantage over the \ac{MILP} can be realized without per-instance training. Second, we will extend the formulation to dynamic request arrivals and to \ac{O-RAN}-specific constraints, such as fronthaul and F1 latency budgets and placement affinities, as well as to multiple slice types with heterogeneous service requirements. Third, we will investigate denser efficiency rewards that provide a stronger learning signal than a terminal bonus.


16KIS2402K
\raggedbottom

\section*{Acknowledgment}
This research was supported in part by the German Federal Ministry of Research, Technology and Space (BMFTR) through the Open6GHub+ project (Grant No.~16KIS2402K) and the 6G Terafactory project (Grant No.~16KISK186).
\vspace{-2mm}


\section*{Acronym List}
\begin{acronym}[Non-RT RIC]
\acro{5G}{fifth-generation}
\acro{6G}{sixth-generation}
\acro{ACO}{ant colony optimization}
\acro{DQN}{deep $Q$-network}
\acro{ILP}{integer linear programming}
\acro{MILP}{mixed-integer linear programming}
\acro{ML}{machine learning}
\acro{NF}{network function}
\acro{NFV}{network functions virtualization}
\acro{Non-RT RIC}{non-real-time RAN intelligent controller}
\acro{O-Cloud}{O-RAN cloud}
\acro{O-CU}{O-RAN central unit}
\acro{O-DU}{O-RAN distributed unit}
\acro{O-gNB}{O-RAN next-generation Node B}
\acro{O-RAN}{open radio access network}
\acro{O-RU}{O-RAN radio unit}
\acro{PM}{physical machine}
\acro{QoS}{quality of service}
\acro{RL}{reinforcement learning}
\acro{SARSA}{state-action-reward-state-action}
\acro{SDN}{software-defined networking}
\acro{SDO}{standards developing organization}
\acro{SFC}{service function chain}
\acro{SLA}{service-level agreement}
\acro{SMO}{service management and orchestration}
\acro{TD}{temporal difference}
\acro{VM}{virtual machine}
\acro{VNF}{virtual network function}
\acro{VNFC}{virtual network function component}
\end{acronym}


\bibliography{mypaper01.bib} 
\bibliographystyle{ieeetr}
\vspace{-1.5mm}


\section*{Author Biographies}
\vspace{-1.5mm}
\begin{IEEEbiography}[{\includegraphics[width=1in,height=1.25in,clip,keepaspectratio]{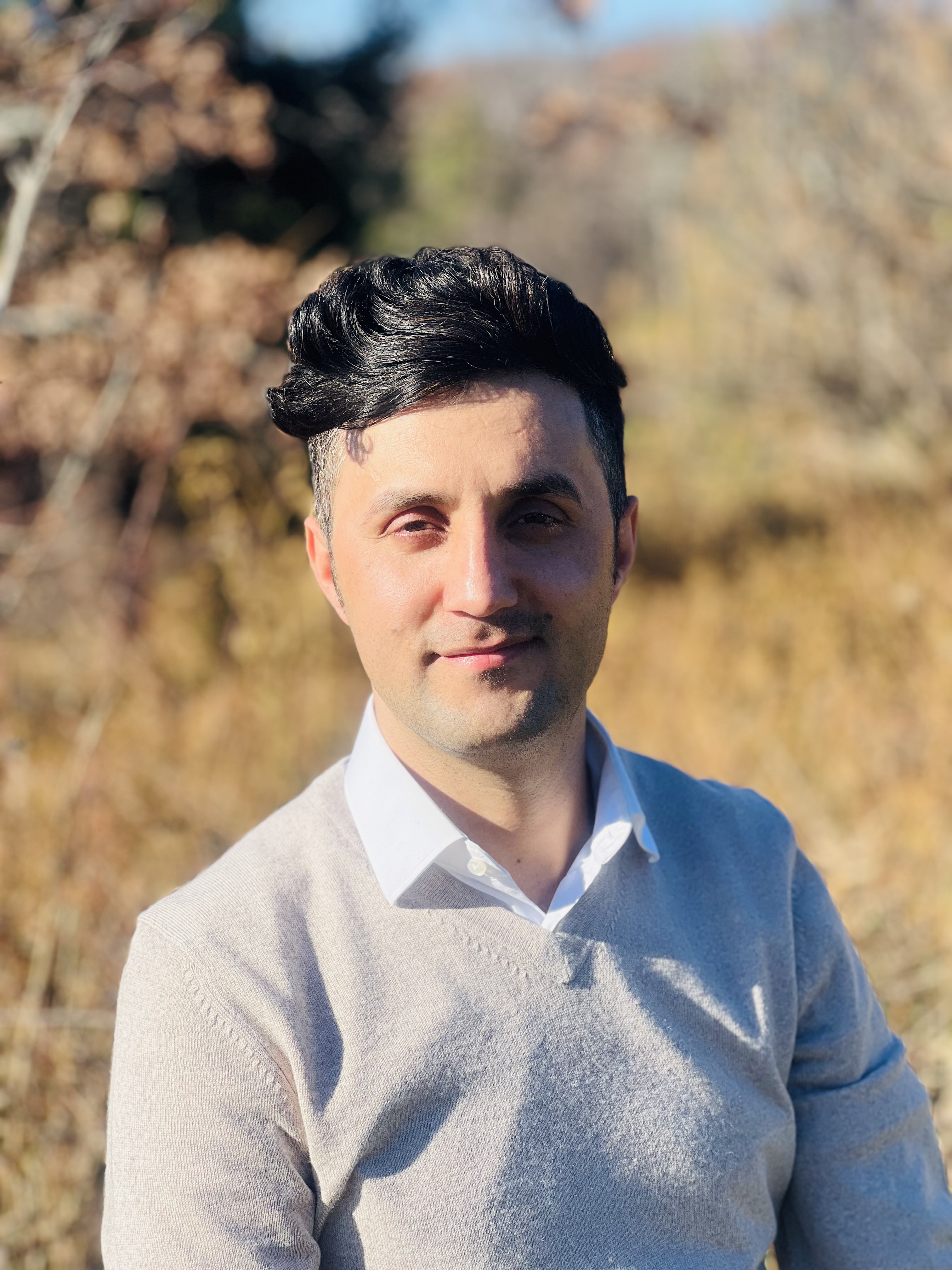}}]{Mohammad Asif Habibi} (M'26) received the B.Sc. degree in telecommunication engineering from Kabul University, Afghanistan, in 2011, the M.Sc. degree in systems engineering and informatics from the Czech University of Life Sciences, Czech Republic, in 2016, and the Dr.-Ing. degree (summa cum laude) from the University of Kaiserslautern (RPTU), Germany, in 2025. From 2011 to 2014, he was a radio access network engineer with Huawei. From 2017 to 2025, he was a Researcher and Senior Researcher with the Division of Wireless Communications and Radio Navigation, University of Kaiserslautern (RPTU). Since 2026, he has been a Senior Researcher with the German Research Center for Artificial Intelligence (DFKI). His research interests include network and service management, network slicing, network architecture, machine learning, and radio access networks.
\end{IEEEbiography}

\begin{IEEEbiography}[{\includegraphics[width=1in,height=1.25in,clip,keepaspectratio]{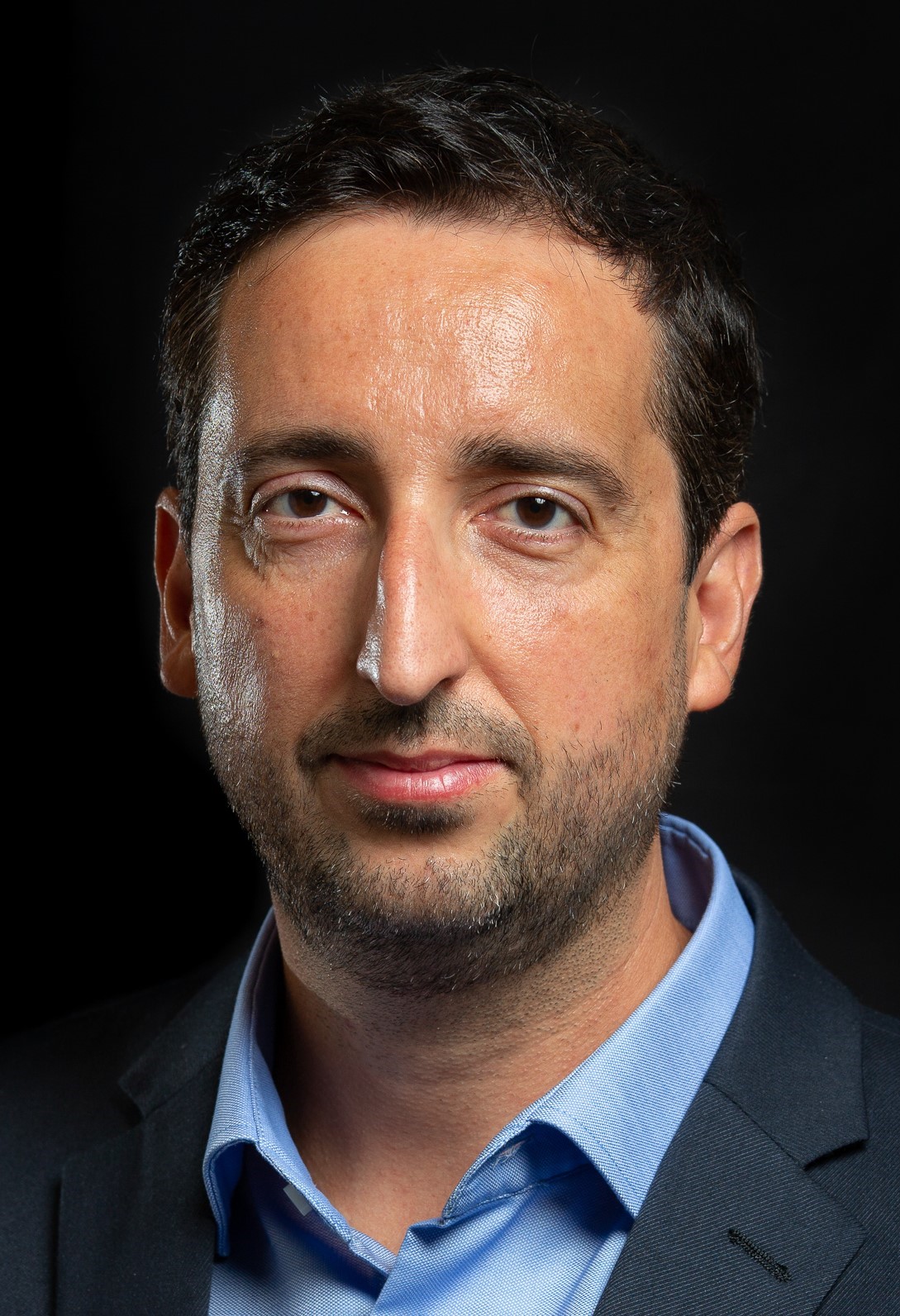}}]{Xavier Costa-P\'erez} (M'06--SM'18) received the M.Sc. and Ph.D. degrees in telecommunications from the Polytechnic University of Catalonia (UPC), Spain, receiving a national award for his doctoral thesis. He is an ICREA Research Professor, Scientific Director at the i2CAT Research Center, and Head of Space and Terrestrial Systems (ASTRO) R\&D at NEC Laboratories Europe. His research focuses on the convergence of space and mobile networks with artificial intelligence. His innovations have been integrated into commercial mobile phones, base stations, and network management systems, and have led to several start-ups. He has contributed to standardization in IEEE 802.11, IEEE 802.16, the Wi-Fi Alliance, and 3GPP, and is an inventor of over 100 granted patents, including standard-essential patents. He has served on the organizing committees of conferences such as AAAI, ACM MobiCom, and IEEE INFOCOM, and as an Editor of the \textsc{IEEE Transactions on Mobile Computing}, the \textsc{IEEE Transactions on Communications}, and \textit{Computer Communications} (Elsevier).
\end{IEEEbiography}

\begin{IEEEbiography}[{\includegraphics[width=1in,height=1.25in,clip,keepaspectratio]{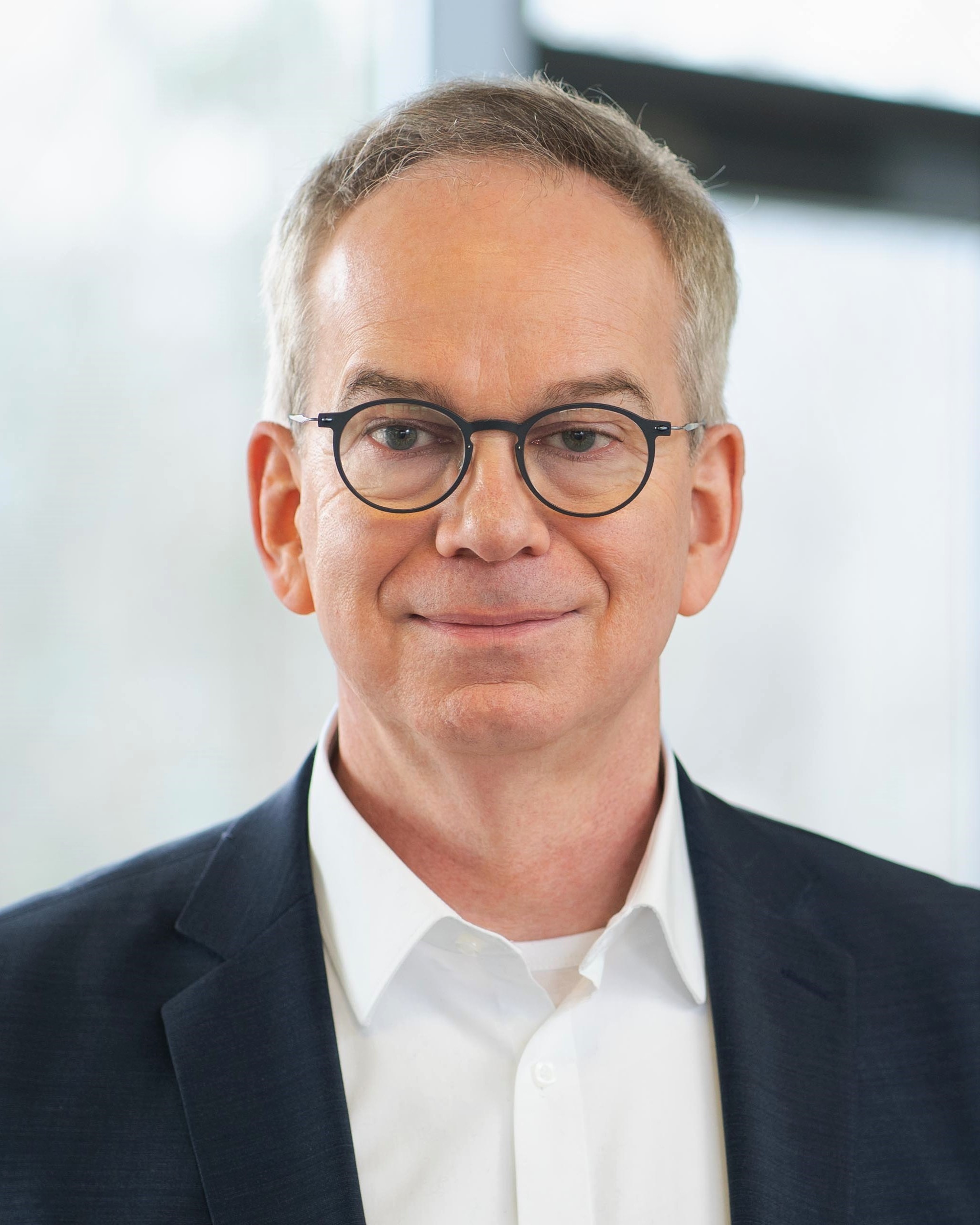}}]{Hans D. Schotten} (S'93--M'97) received the Diploma and Ph.D. degrees in electrical engineering from the Aachen University of Technology, Germany, in 1990 and 1997, respectively. Since August 2007, he has been a Full Professor and Head of the Division of Wireless Communications and Radio Navigation, University of Kaiserslautern (RPTU). Since 2012, he has also been Scientific Director at the German Research Center for Artificial Intelligence (DFKI), heading the Intelligent Networks Research Department. Previously, he held positions as a senior researcher, project manager, and research group head at Aachen University of Technology, Ericsson Corporate Research, and Qualcomm Corporate R\&D, where he was also Director for Technical Standards and coordinated Qualcomm's activities in European research programs.
\end{IEEEbiography} 

\end{document}